\documentclass{article}
\usepackage[utf8]{inputenc}

\usepackage{authblk}

\title{Cusp bifurcation in a metastatic regulatory network}
\author[1]{Brenda Delamonica}
\author[2]{G\'abor Bal\'azsi$^\ast$}
\author[3]{Michael Shub$^{\ast \ast}$}
\affil[1]{Applied Mathematics and Statistics Department, Stony Brook University, Stony Brook, NY 11794, USA}
\affil[2]{The Louis and Beatrice Laufer Center for Physical and Quantitative Biology, Stony Brook University, Stony Brook, NY 11794, USA \\ Department of Biomedical Engineering Department, Stony Brook University, Stony Brook, NY 11794, USA}
\affil[3]{Department of Mathematics,City College and the Graduate Center of CUNY}
\date{}
\usepackage{color,soul}
\usepackage{graphicx}
\usepackage{amsmath}
\usepackage{amssymb}
\usepackage{caption}
\usepackage{subcaption}
\usepackage{float}
\usepackage{wrapfig}
\usepackage{parskip}
\usepackage{indentfirst}
\usepackage[english]{babel}
\usepackage[a4paper, total={6in, 8in}]{geometry}
\PassOptionsToPackage{hyphens}{url}
\usepackage[breaklinks]{hyperref}
\usepackage[dvipsnames]{xcolor}
\begin{document}

\maketitle
$\ast, \ast \ast
$ Corresponding authors: gabor.balazsi@stonybrook.edu, shub.michael@gmail.com

Declaration of Interest: None.

\newpage

\tableofcontents

\newpage

\begin{abstract}
{Understanding the potential for cancers to metastasize is still relatively unknown. While many predictive methods may use deep learning or stochastic processes, we highlight a long standing mathematical concept that may be useful for modeling metastatic breast cancer systems. Ordinary differential equations (ODEs) can model cell state transitions by considering the pertinent environmental variables as well as the paths systems take over time. Bifurcation theory is a branch of dynamical systems which studies changes in the behavior of an ODE system while one or more parameters are varied. Many studies have applied  concepts in one-parameter bifurcation theory to model biological network dynamics, and cell division. However, studies of two-parameter bifurcations are much more rare. Two-parameter bifurcations have not been studied in metastatic systems. Here we show how a specific two-parameter bifurcation phenomenon called a cusp bifurcation separates two qualitatively different metastatic cell state transitions modalities and propose a new perspective on defining such transitions based on mathematical theory. We hope the observations and verification methods detailed here may help in the understanding of metastatic potential from a basic biological perspective and in clinical settings.}
\end{abstract}
\newpage
\section{Introduction}

Different cell states can emerge during disease progression, such as cancer metastasis. Regarding metastatic cancer, much attention has been devoted to two cellular states: epithelial (E) and mesenchymal (M), each recognizable by the levels of specific proteins, which correspond to the steady states of multistable gene regulatory networks \cite{PMID:19487818, PMID:10647931,PMID:33500360,PMID:16493418,PMID:27840647}. Normal epithelial (E) cells are not motile and can grow (divide) in response to growth signals, as opposed to mesenchymal (M) cells that do not form epithelial layers and are motile, but less likely to divide \cite{PMID:27634432}. The assumption of a binary choice between cell division and movement is the ``go-or-grow hypothesis" \cite{PMID:18245463}. Since metastasis requires departure from a primary site and growth at a different site, both the epithelial-mesenchymal transition (EMT), and mesenchymal-epithelial transition (MET) seem to be required \cite{PMID:23238008}. Thus, EMT alone is not always sufficient for metastasis \cite{PMID:26560033} if a binary classification into E and M cell types is assumed since MET is also required for growth in the new location. Moreover, EMT might not be necessary either, according to recent work \cite{PMID:26560033} \cite{PMID:26560028} showing that most metastatic cells do not undergo full EMT, and EMT inhibition does not reduce metastasis. Furthermore, metastasis relies on detached cells invading their tissue neighborhood and accessing the bloodstream, which occurs on top of EMT, under the control of different genes called pro- and antimetastatic regulators, such as BACH1 and RKIP \cite{metarticle}. We believe that we can pinpoint the threshold when these regulators cause a qualitative shift in the transition between the two states, and describe a new view that may be crucial to understanding whether and how metastasis will occur.

Many recent studies indicate that EMT and MET are more complex than binary processes, i.e., they are transitions between more than two distinct, well-defined cellular states \cite{PMID:25270257,PMID:28548345,PMID:24154725}. One or more intermediate, ``hybrid" or ``partial" EMT cell states with mixed E/M properties have been described \cite{PMID:29920274,PMID:32217696,PMID:29670281}. Recent computational work that varied the number of hidden intermediate states, aiming to improve fits to experimental data \cite{PMID:32155144} found that intermediate states can accelerate EMT. It seems possible that only these intermediate EMT states, instead of full EMT, are necessary and sufficient for metastasis \cite{PMID:34115987}.  In general, the number of such intermediate states is unknown, raising the question: Can the number of intermediate states go to infinity, allowing continuous transitions? And what defines the boundary between such continuous transitions versus the widely-studied discrete EMT and MET transitions, with a finite number of distinguishable steady states? Finally, how can these theoretical questions help us to understand the biology of metastasis?

The use of equilibrium states of differential equations as a model for biological or chemical systems including metastasis-regulatory networks, has a long history which we do not try to survey here, except for a few instances. \cite{recref1,recref2,ref3thom1,ref4thom2,ref5,ref6toggle,ref7toggle,ref8rep} The study of bistable systems has played an important role, especially noteworthy are the works on the toggle  switch \cite{ref6toggle,ref7toggle} and on direct and indirect fully positive feedback \cite{11350942} \cite{PMID:11891111}. (A deeper example of these models and how the relate to cusp bifurcations can be found in the Supplementary Materials Appendix A)

Mathematical studies of genetic regulatory networks have usually relied on solving the corresponding differential equations numerically or with the use of topological or fixed point techniques and theorems from one-parameter bifurcation theory to prove the existence of solutions with particular properties. One example of the use of topological methods describing critical points and associated Boolean networks is by Glass \cite{Glass}. An example of the extensive use of numerics is Lee et al \cite{metarticle}. By contrast, the use of two-parameter bifurcation theory rarely if ever enters into the biological theory, even in papers with the word bifurcation in the title as in Spencer et al. \cite{spenceretal} or in the text as in Rajapakse et al. \cite{rajapakse2016mathematics} The cusp bifurcation is a concept from two-parameter bifurcation theory, found by solving a system of equations in the state and parameter variables. Here we present methods for finding cusp points and give a self-contained elementary derivation of a set of equations and their solution for finding cusp points. We provide below a simple computational framework to find accurate solutions to such systems. We show how this provides a broader view on the emergence of bistability in biological systems, by dividing the two-dimensional parameter space into regions with distinct transition types: continuous and discontinuous transitions. This should improve the conceptual understanding of cell state transitions in metastatic gene networks and other biological systems.

\section{The cusp point separates two kinds of cell state transitions}

Cell states or ``cell types" \cite{rajapakse2016mathematics} have been modeled as stable steady states, or equilibria, of ordinary differential equations $$\dot{x}=V(x,\alpha)$$ such that $x \in \mathbb{R} ^n$ is a concentration vector representing the cell's molecular composition and $\alpha \in \mathbb{R}^j$ is a vector of $j$ parameters, representing internal and external characteristics, including reaction rates. Steady states correspond to the values of $x_\alpha$ which satisfy $V(x_\alpha,\alpha)=0$. If the eigenvalues of the derivative of $V(x_\alpha,\alpha)$, denoted $D_x V(x_\alpha,\alpha)$, all have negative real part then nearby solutions all tend to $x_\alpha$ as time increases. It is possible that $V(x,\alpha)$ may have a unique stable state, multiple stable states or even more complicated dynamical behavior. Each stable equilibrium $x_{\alpha,i}$ corresponds to a cell type. If the parameter $\alpha$ depends on some external factor $f_{ext}$ because of sensory, genetic, epigenetic, spatial or other effects, then the number and value of stable states $x_{\alpha(f_{ext})}$ may depend on $\alpha(f_{ext})$, which describes the dynamical behavior of cells transitioning from one type to another dependent on $f_{ext}$, which could be the time variable.  We are interested in studying the transitions which involve pro-metastatic and anti-metastatic mono-stable states of cells, analogous to EMT and MET, using mathematical models derived from bifurcation theory. In the model $\dot{x} = V(x,\alpha)$, the variables $x(t)$ that represent proteins involved in metastatic cell transitions are time-dependent. The parameters $\alpha(f_{ext})$ depend on an external factor, which could be a chemical concentration, cell size, or time. As long as $\alpha(f_ext)$ is a continuous function of $f_ext$, the theory we present is valid. If the factor $f_ext = t$ is time, as we assume in the following for simplicity, the adjustment time scale of the steady state would be faster than the time scale of $\alpha(t)$, as usually assumed in bifurcation theory.
We will mainly focus on cell types in metastatic breast cancer, but such analyses may also be generalized to other metastatic cancers, biological networks orchestrating events such as cell division, \cite{spenceretal} or synthetic gene circuits \cite{rajapakse2016mathematics}.

Smale and Rajapakse refer to cell states as ``cell types" in  \cite{rajapakse2016mathematics} where they identify conditions in biological networks for which a pitchfork bifurcation in 2 and 3 variable systems with one parameter exist. Yet, the pitchfork is a one-parameter bifurcation that may not be stable. There is a stable two parameter bifurcation called the cusp bifurcation, which includes a pitchfork as a one dimensional sub-bifurcation. The applicability of the cusp bifurcation or two-parameter bifurcations to metastatic transitions has not been widely investigated. Here we use known methods \cite{Pujals2018StableAN} to verify that a pitchfork exists in the metastasis model by Lee et al.\cite{metarticle} (See sections 2.2 and 2.3). Moreover it can be shown that all the examples of pitchfork bifurcations proven by Smale and Rajapakse in \cite{rajapakse2016mathematics} concerning ``Repressillator" and ``Toggle" synthetic gene circuits are actually one dimensional sub-bifurcations of cusp bifurcations (See Supplementary Materials). We plot the cusp curve in the metastatic breast cancer model, which is the projection of the fold onto the parameter space and observe that it divides the parameter space into two regions separated by the curve and cusp point, which correspond to biological transitions of two types. These two transition types correspond to the two types of paths taken by the curve $\alpha(t)$. The first goes around the the cusp point, and has no bistable points, whereas the second crosses the bistable region. The first we call a continuous transition, as it may capture the biological phenomenon of not just one, but any number of ``hybrid" or ``partial" cell types, differently than in previous studies. Specifically, instead of multiple equilibria, we find that there is always only one stable equilibrium in the dynamical system, corresponding to a continuum of partial cell types, which may be sufficient for metastasis. The second transition is a discontinuous transition between binary cell types, or anti-metastatic and pro-metastatic cells, which happens when a stable equilibrium bifurcates and two stable equilibria and one unstable equilibrium (saddle) appear. 

Overall, we show a novel application of bifurcation theory in biology, propose a shift from continuous (non-binary) to discrete (binary) transitions at the cusp point, and discuss further applications of these concepts in metastasis and other biological phenomena. We have also provided a tutorial in bifurcation theory and any relevant code in the Supplementary Materials, which others may find useful for their own research.

\section{Cusp bifurcation for metastatic cell state transitions}

\subsection{Analytical methods for finding the cusp point: The metastatic cell transition ODE model}

First, we derive mathematical conditions that can identify a cusp bifurcation using the model from the paper “Network of mutually repressive metastasis regulators can promote cell heterogeneity and metastatic transition"\cite{metarticle}. Here, Lee et al. considered three differential equations with two parameters $V(R,L,B,\rho,k)$ where $R, L, B$ are real positive variables and $\rho, k$ are real positive parameters. $R, L, B$ represent the proteins and RNAs RKIP, let-7 and BACH1 which interact in the cell and are highly relevant for determining breast cancer metastasis. The parameters $\rho$ and $k$ describe the instability of RKIP and insensitivity of BACH1 to self-regulation, respectively. The equations are
\begin{align*}
\frac{dR}{dt} & = \frac{1}{1 + B} - \rho R & \\
\frac{dL}{dt} & = \frac{a R^r}{m^r + R^r}-L-cLB & \equiv V(R,L,B,\rho,k) \\
\frac{dB}{dt} & = s +\frac{(S-s)k^b}{k^b + B^b} - B - cLB &
\end{align*}
The constants are set to $s=.02, S=20, c=200, m=2, b=3, r=5, p=10, a=1000$. Let $\Vec{x}= (R, L, B)$ so for convenience we may write $V(\Vec{x},\rho,k)$ and take derivatives with respect to $\Vec{x}$. 

Observable  cell states are equilibria of the system $V(\vec{x},\rho,k)$. For $\rho, k$ fixed, equilibria are points $\Vec{x}$ such that $V(\Vec{x},\rho,k)=0$. The equilibrium is stable if the real parts of the eigenvalues of $D_x V(\Vec{x},\rho,k)$ are negative. All solutions of the ODE which start near a stable equilibrium tend to the equilibrium as time increases. An equilibrium may lose its stability and a bifurcation may occur as we vary $(\rho, k)$ if one of the eigenvalues tends to have zero real part or more specifically, if the eigenvalue becomes zero. In other words, to find bifurcations, we are looking for the solutions of the determinant $Det[D_x V(\Vec{x},\rho,K)]=0$. 
 
\begin{figure}[H]
\begin{subfigure}{0.5\textwidth}
\centering
\includegraphics[scale=.6]{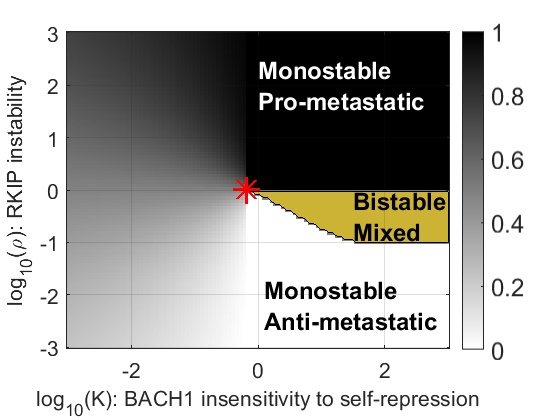}
\caption{}
\label{fig:1a}
\end{subfigure}
\begin{subfigure}{0.5\textwidth}
\includegraphics[scale=.3]{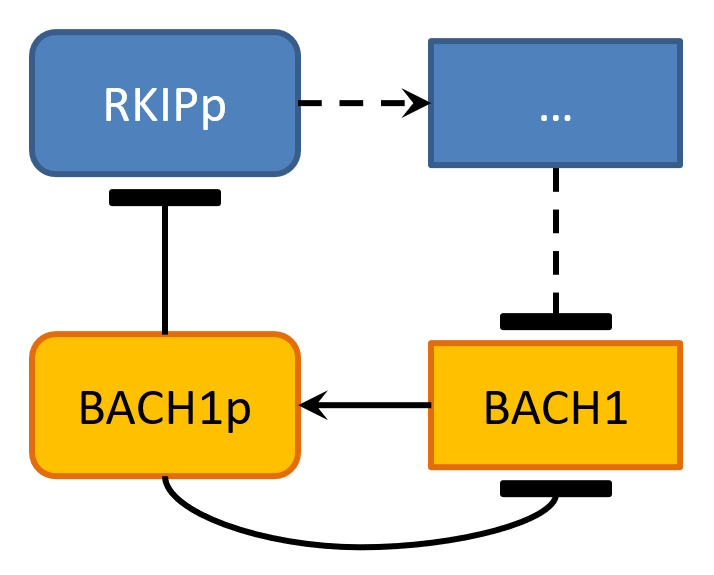}
\centering
\caption{}
\label{fig:1b}
\end{subfigure}
\caption{ (a)Map of dynamics from Lee et al. \cite{metarticle}. Grayscale shading indicates monostability. Darker shading corresponds to higher BACH1 levels. Yellow color indicates bistabiltiy. The red star indicates the expected cusp point where ($\rho$,K) = (1.0024,0.6595). (b) Gene network diagram of BACH1 and RKIP regulatory interactions. BACH1 is BACH1 gene, BACH1p is BACH1 protein; RKIPp is RKIP protein. The dots represent intermediary regulators between RKIP and BACH1.} 
\label{fig:1}
\end{figure}

Via an intricate analysis of the ODE, Lee et al. divide a region in the $(\rho,k)$ parameter plane into three sub-regions (Figure \ref{fig:1}). One region with a single stable equilibrium corresponding to an anti-metastatic state of the cell, one with a single stable equilibrium corresponding to a pro-metastatic state of the cell and one with three equilibria, two of which are stable. This suggests that as $\rho,k$ vary the state of the cell may start in one monostable region and pass through a bistable region to the other monostable region. Thus the boundary separating the bistable and mono-stable regions is a curve of interest. It is defined by the solution of
\begin{align*}
& V(\Vec{x},\rho,k)=0 \\
& Det[D_x V(\Vec{x},\rho,k)]=0
\end{align*}
There are now 4 equations in 5 unknowns to solve. We assume that the rank of the derivative of this system is 4 when the equations are satisfied. By the implicit function theorem we may graph a curve for an underdetermined system \cite{pugh} to (locally) locate the set of solutions. The curve projected onto the $(\rho,k)$-plane is smooth and locally separates the regions of the plane. Under certain conditions this curve may meet at a cusp point (See Appendix B). The shape of the bistable region is such that we suspected that there is a cusp point, which we do in fact find to be close to the red star in Figure \ref{fig:1}.

A cusp point is the $(\rho,k)$ parameter coordinates of a non-degenerate solution of the following five by five system of equations.
\begin{equation}
    \begin{split}
    V(\Vec{x},\rho,k) &=0 \\
    Det[D_x V(\Vec{x},\rho,k)] &=0 \\
    \nabla_x Det[D_x V(\Vec{x},\rho,k)]\bullet (\Vec{v})&=0
    \end{split}
\end{equation}

Here $\nabla_x$ denotes the gradient in terms of $\vec{x}$ which we take of the determinant of the derivative. We then calculate the dot product with $(\Vec{v})$ which is the first column of the adjugate matrix of $D_x V(\Vec{x},\rho,k)$. We assume $(\Vec{v})$ is not zero. The rank of $D_x V(\Vec{x},\rho,k)$ is two where the equations are satisfied  because $D_x V(\Vec{x},\rho,k)$ is a 3 by 3 matrix with an eigenvalue equal to zero. Since the map has maximal rank it is stable even if the parameters vary slightly. The solution to the equations (1) define the values of proteins $R,L,B$ and the parameter values $\rho, k$ required at the cusp point.

\begin{figure}[H]
\begin{subfigure}{0.5\textwidth}
\includegraphics[scale=.6]{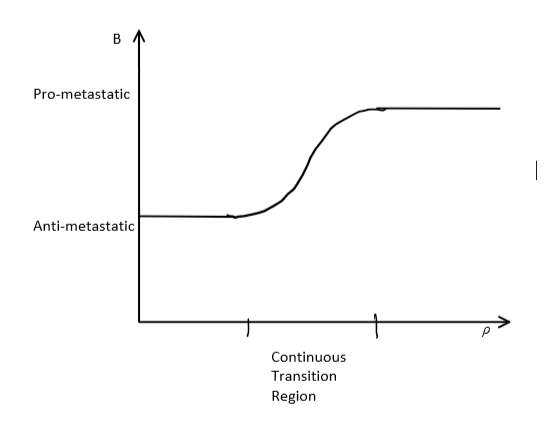}
\caption{Path of $\alpha$ outside of bistable region}
\label{fig:2}
\end{subfigure}
\begin{subfigure}{0.5\textwidth}
\includegraphics[scale=.6]{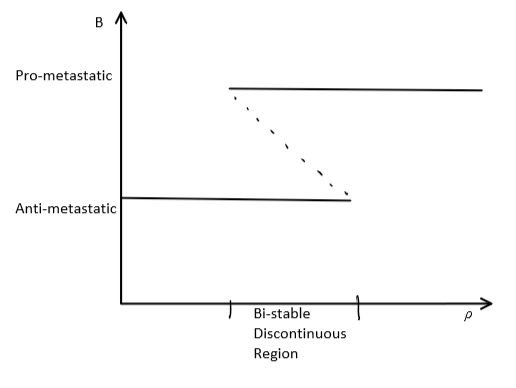}
\caption{Path of $\alpha$ through bistable region}
\label{fig:3}
\end{subfigure}
\caption{Two different modes of cell state transitions between monostable regions. Image (a) demonstrates a continuous transition cell state path and (b) desmonstrates a bistable discontinuous transition cell state path.}
\label{fig:4}
\end{figure}

In order to observe metastasis, cells must traverse from the bottom to the top and back in Figure \ref{fig:1}. To the left of the cusp, which is the grey area of Figure \ref{fig:1}, we have cells that can transition continuously from one state to another (Figure \ref{fig:2}). To the right of the cusp, cells must cross the bistable region in yellow in the upward direction and then again in the downward direction, where the cell plots are discontinuous (Figure \ref{fig:3}). We explain this further in the Discussion section below. Mathematically, these paths can be observed by finding solutions of the system where either $k$ or $\rho$ are fixed. In the first case, $k$ is fixed and must be less than its cusp point value, and in the second $k$ must become greater than the cusp point value. In Figure \ref{fig:4} we draw sketches of what we expect these solutions to look like and how they relate to metastasis. 

\subsection{Numerical approach for finding the cusp point and model verification}

We built a Newton’s Method algorithm in MATLAB for an underdetermined system of equations to validate the $\rho,k$ solution values we identified from Lee et al. We used MATCONT\cite{matcont}, a continuation toolbox for ODEs, to isolate a more accurate set of values for the cusp point on the $(\rho,k)$-plane and plot the projection of the cusp curve. We also built codes in MATLAB to visualize the various behaviors of the system which can be found in the supplementary materials.

\begin{figure}[H]
\begin{subfigure}{0.5\textwidth}
\includegraphics[scale=.5]{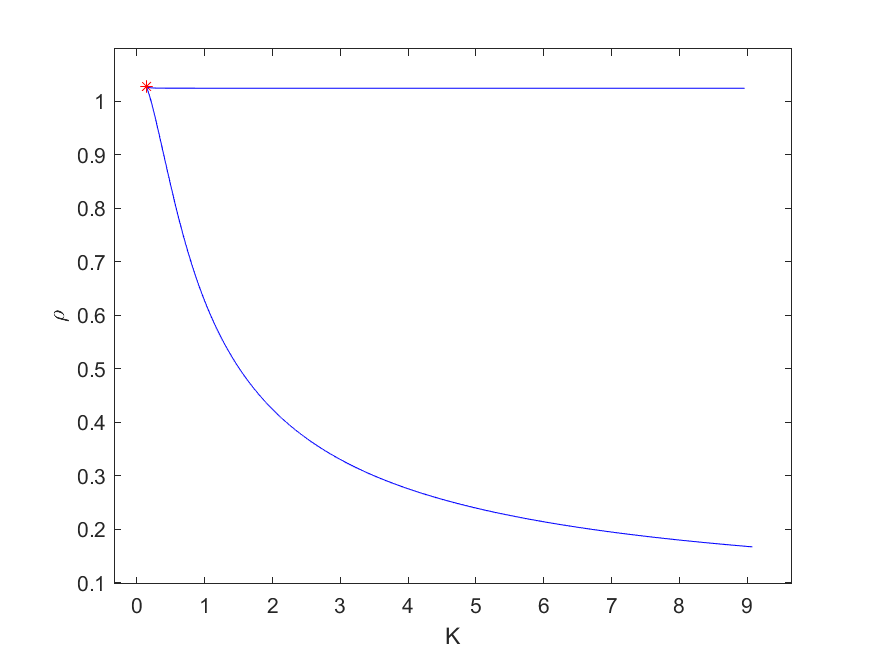}
\caption{MATCONT plot of the cusp bifurcation}
\label{fig:6}
\end{subfigure}
\begin{subfigure}{0.5\textwidth}
\includegraphics[scale=.5]{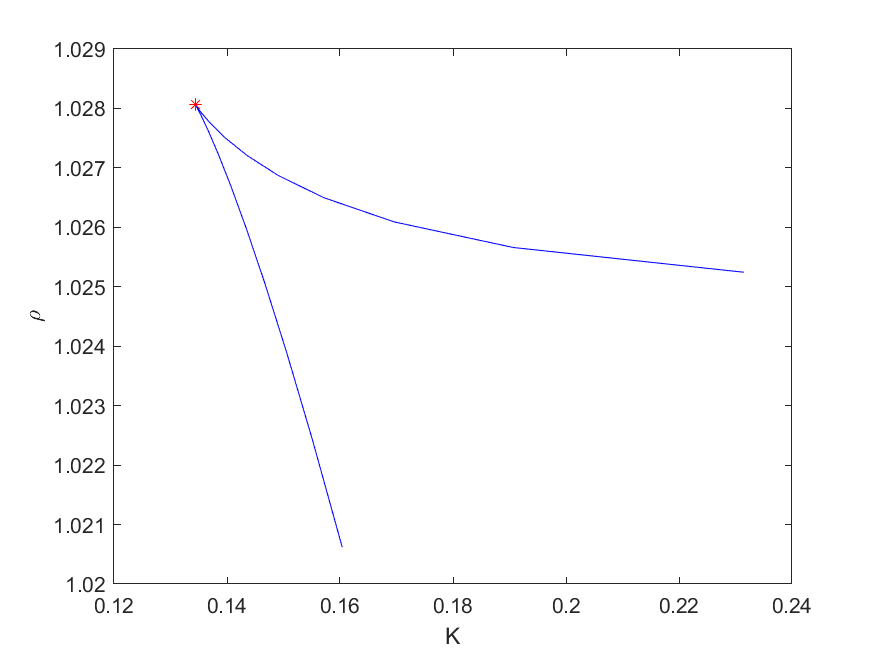}
\caption{Newtons Method solution curves around the cusp}
\label{fig:7}
\end{subfigure}
\caption{The cusp point plot in (a) is the output from MATCONT when running the Limit Cycle differential equation solver for a Cusp Point. In plot (b) we used our Newton's method algorithm to verify numerically the same plane division using the equations from Lee et al.The diamond is the MATCONT generated values of $(\rho,k)=(1.0281, 0.1343) $}
\label{fig:8}
\end{figure}

To validate this is in fact a cusp point, we solved the system of equations in MATCONT to find that the cusp is located at $(\Vec{x},\rho,k) = (0.9321, 2.2184, 0.0435, 1.0281, 0.1343)$.  Note that the values from MATCONT return $\rho,k=( 1.0281, 0.1343)$ whereas the previous values in Figure \ref{fig:1} were $(1.0024,0.6595)$. When plotting $\rho,k$ we found a similar separation of the regions defined by Lee et al. (\ref{fig:8}).

%\begin{figure}[H]
%\centering
%\includegraphics[scale=.7]{NewtonsMethodresults.JPG}
%\caption{Results using Newton’s Method algorithm}
%\label{fig:5}
%\end{figure}

Next, we numerically verified that the five by five system has an invertible derivative at the solution. At a cusp point the tangent to the cusp exhibits a pitchfork bifurcation. \cite{Pujals2018StableAN} 
 
\begin{figure}[h]
\centering
\includegraphics[scale=.7]{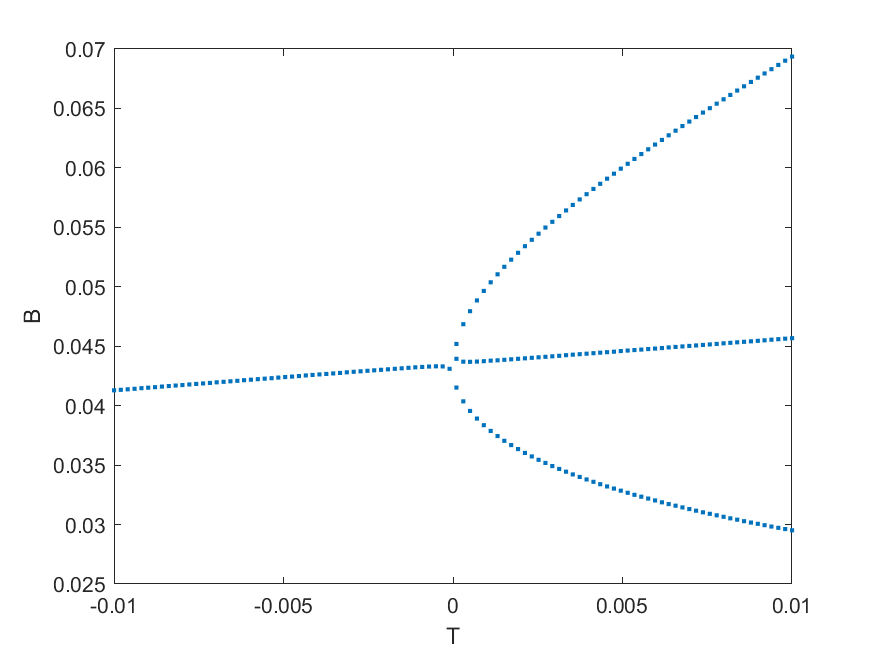}
\caption{Pitchfork Bifurcation plot. The solution of $B$ is derived around the new values of $(\rho, k)=(1.028071,0.134353)$ using the Matlab function vpasolve().}
\label{fig:9}
\end{figure}
 
Figure \ref{fig:9} is a plot of the pitchfork bifurcation around the cusp point $(\rho,k) = (1.0281,0.1343)$ where the tangent vector is (-0.1542,1). For values of $T$ moving in the positive direction of the tangent, $B$ consistently has 3 solutions, two stable equilibria which are the upper and lower branches of the pitchfork and one unstable equilibria in between. In the negative direction of $T$ we find only 1 solution.

\begin{figure}[H]
\begin{subfigure}{0.5\textwidth}
\includegraphics[scale=.5]{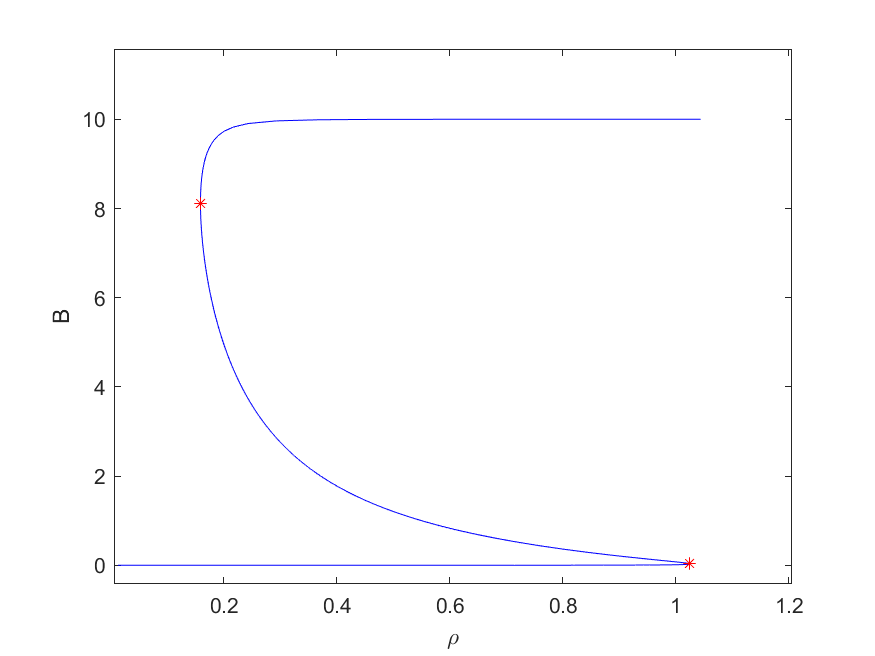}
\caption{k=10}
\label{fig:10}
\end{subfigure}
\begin{subfigure}{0.5\textwidth}
\includegraphics[scale=.5]{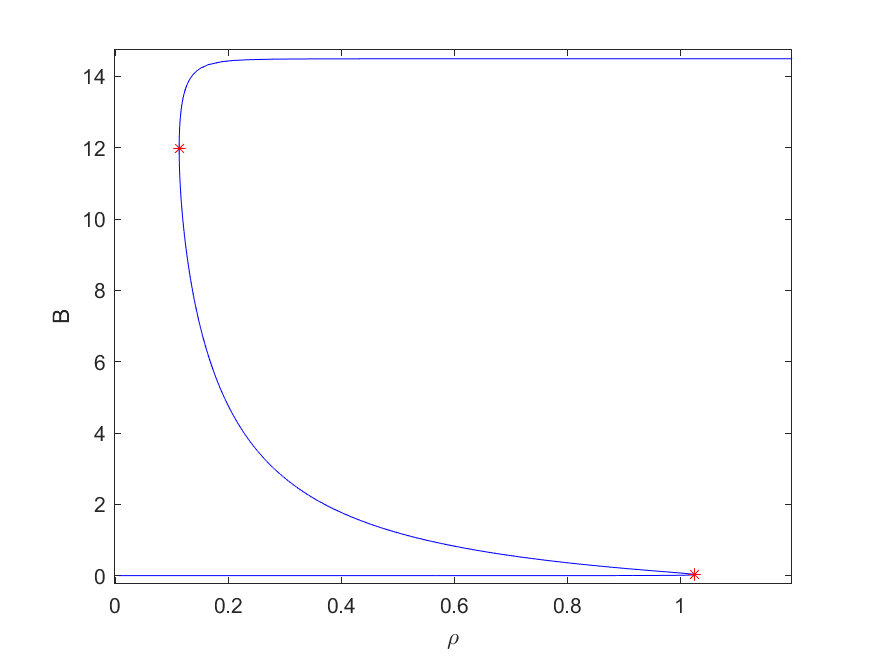}
\caption{k=20}
\label{fig:11}
\end{subfigure}
\caption{(a) is a solution curve from MATCONT's Equilibrium curve solver where $k=10$ was fixed and $\rho$ varied. (b) is a solution curve from MATCONT's Equilibrium curve solver where $k=20$ was fixed and $\rho$ varied. In both plots the red stars are limit points, desmonstrating discontinuous solutions to the system.}
\label{fig:12}
\end{figure}

We explore how varying values of $\rho$ affects $B$ in Figure \ref{fig:12}. First we note that near the cusp point for different values of $k > .13$ we see a similar curve. If we follow a solution from the lower monostable region it disappears as $\rho$ increases, at 1.024. This is the first limit point (LP) or equilibrium in the graph. Then $B$ increases towards the value corresponding to the other stable equilibrium which is the upper limit point. The curve between the two limit points represents the bistable region which we showed in Figure \ref{fig:3}. The bistable region in Figure \ref{fig:1} corresponds to the unstable equilibrium and the space between the two limit points. At the second limit point $B$ crosses from the bistable to the monostable region. As $\rho$ increase $B$ stays in the upper branch of the curve. As $\rho$ is approaching zero $B$ tends towards the upper limit point. The limit points are where $\rho$ crosses the cusp line for fixed $k$, thus demonstrating hysteresis. We may interpret this as having a concentration of antimetastatic cells for values of $\rho =\{0,1.024\}$ and a concentration of prometastatic cells from $\rho  = \{.761, \infty\}$ which ``mix" in the bistable region.

\begin{figure}[H]
\centering
\includegraphics[scale=.6]{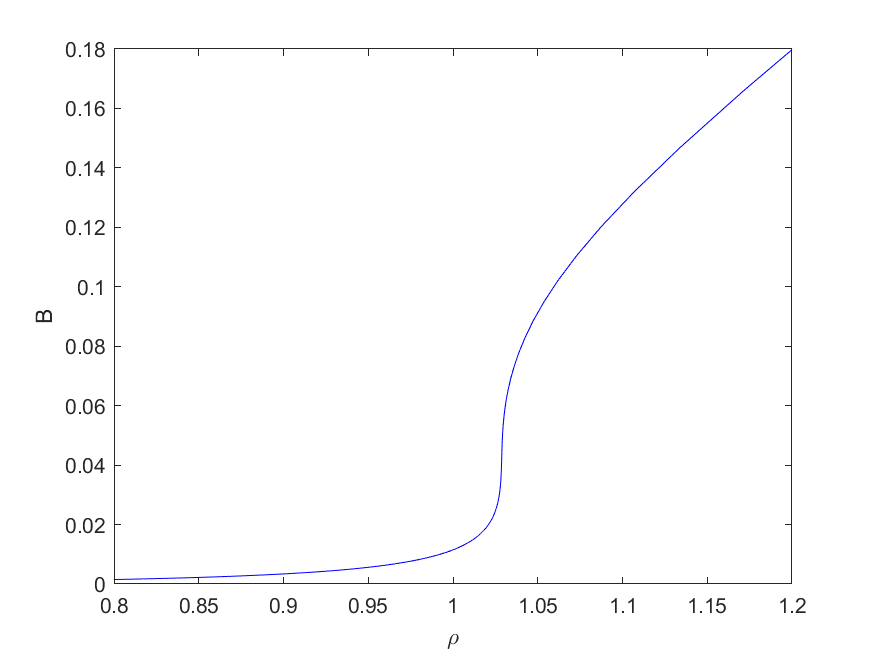}
\caption{A continuous solution curve plot from MATCONT's Equilibrium curve differential equations solver. This image is meant to demonstrate how varying $\rho$ for a fixed value of $k< .13$ will yield a continuous uninterrupted curve without any cusp or limit points.}
\label{fig:13}
\end{figure}

In Figure \ref{fig:13}, if $k = .13$ or less than the value required at the cusp, the solution curve is S shaped and $B$ is increasing continuously as $\rho$ increases. As we continuously increase $\rho$ we pass from the anti-metastatic to pro-metastatic cell states as in Figure \ref{fig:3}. In this scenario, if points on the solution curve correspond to various cell types, we could posit that we transition to many intermediate cell states as we go from one mono-stable state to another. This differs from the case when $k > .13$ where we see two distinct sets of cell types.

\begin{figure}[H]
\begin{subfigure}{0.5\textwidth}
\centering
\includegraphics[scale=.5]{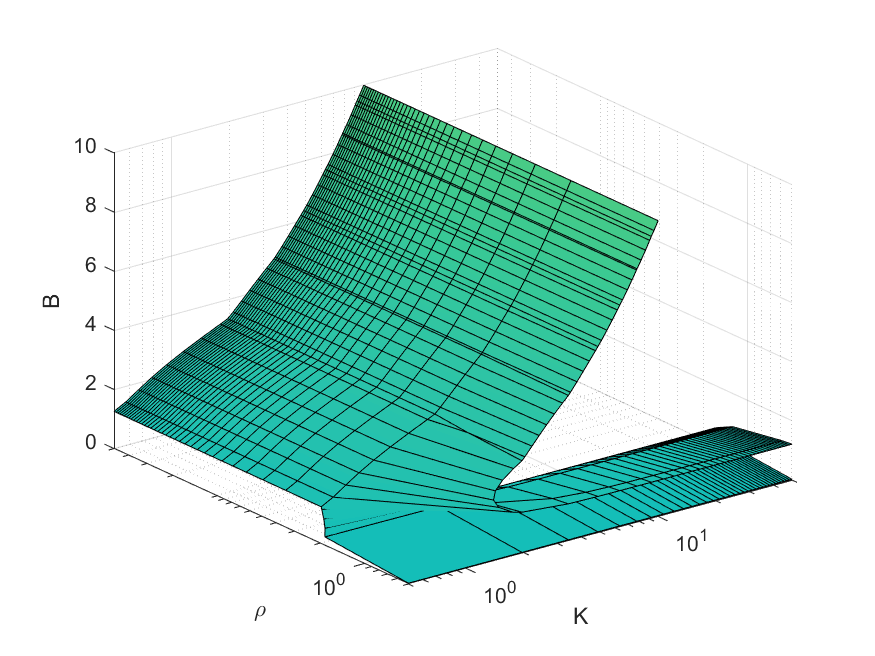}
\caption{}    
\end{subfigure}
\begin{subfigure}{0.5\textwidth}
\centering
\includegraphics[scale=.5]{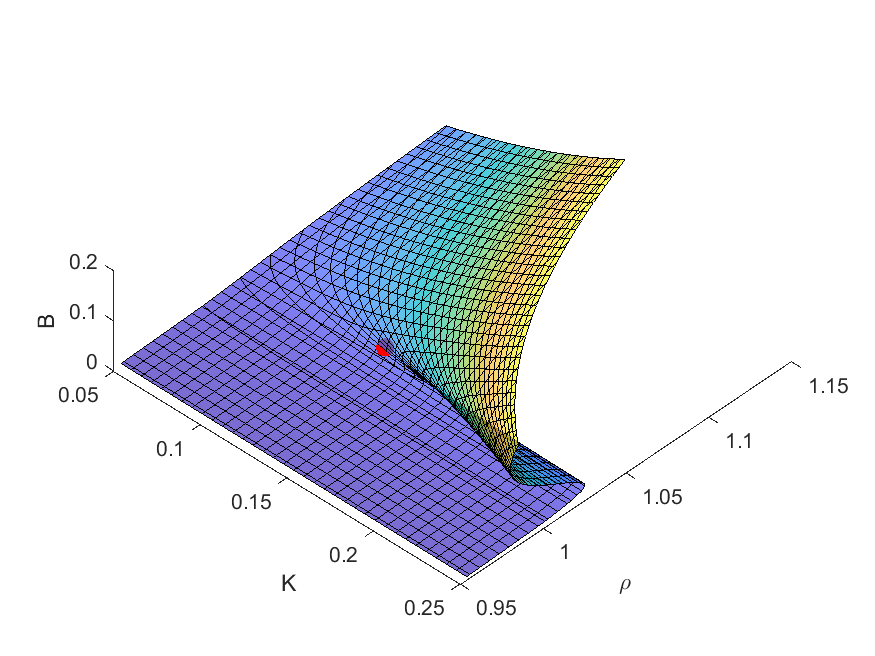}
\caption{}    
\end{subfigure}
\caption{(a) Cusp fold plot of $\rho$ and k using the Matlab function fimplicit3(). At fixed values of $k$ we see $\rho$ varies continuously, transitioning from regions above and below the fold. At the cusp point the surface intersection along fixed $k$ is a pitchfork. The projection of the surface around the cusp onto the $\rho,k$ would yield the cusp point plots above. (b) Cusp fold image zoomed in near the cusp point in red.}
\label{fig:14}
\end{figure}

Finally, in Figure \ref{fig:14} we plot the surface $B$. If we intersect the solution plane at a fixed value of $k$ the solution curves are similar to Figure \ref{fig:12} depending on the choice of $k$. If we intersect the plane at the cusp point where the surface folds we have a solution curve which looks like a pitchfork, as demonstrated in Figure \ref{fig:9}. Furthermore, if we project the fold lines of the surface onto the $\rho, k$ plane we end up with a curve and cusp point as seen in Figure \ref{fig:13}.

%\begin{figure}[H]
%\centering
%\includegraphics[scale=.5]{final_3.PNG}
%\caption{Cusp fold plot analysis}
%\label{fig:15}
%\end{figure}

\section{Discussion}

Many attempts to predict metastasis have been made using a variety of clinical methods and computational techniques \cite{metmap} \cite{metanet} \cite{metabb}. One great challenge is understanding how the environment of early stages of cancer may determine metastasis later on. None of the earlier studies of one-parameter EMT regulatory network dynamics incorporate the information we have found regarding cusp bifurcations in metastatic systems, which we model differently from the earlier EMT studies, by using two-parameter bifurcation theory to analyze a different gene network directly involved in metastasis. Since the theory of two-parameter bifurcations and cusp bifurcation analysis is generally applicable, we believe these findings may complement one-parameter bifurcation studies on EMT and enhance existing attempts to develop predictive methods. In our study we have identified a cusp bifurcation in the model provided by Lee et al. which suggests to us that an interesting behavior, similarly important as hysteresis is for one-parameter saddle-node bifurcatons, may occur in metastasis more generally, where the path of the system will determine final metastatic states. It may even be possible to find this behavior in other cancers or caner-related phenomena governed by bistable or multistable regulatory networks. Below we describe one possible interpretation of how the cusp may determine certain distinct system pathways towards metastasis.

\begin{figure}[H]
\begin{subfigure}{0.4\textwidth}
\centering
\includegraphics[scale=.5]{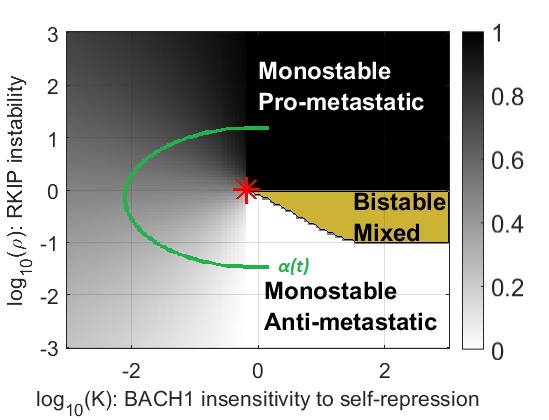}
\caption{ }
\label{fig:16a}
\end{subfigure}
\begin{subfigure}{0.4\textwidth}
\centering
\includegraphics[scale=.2]{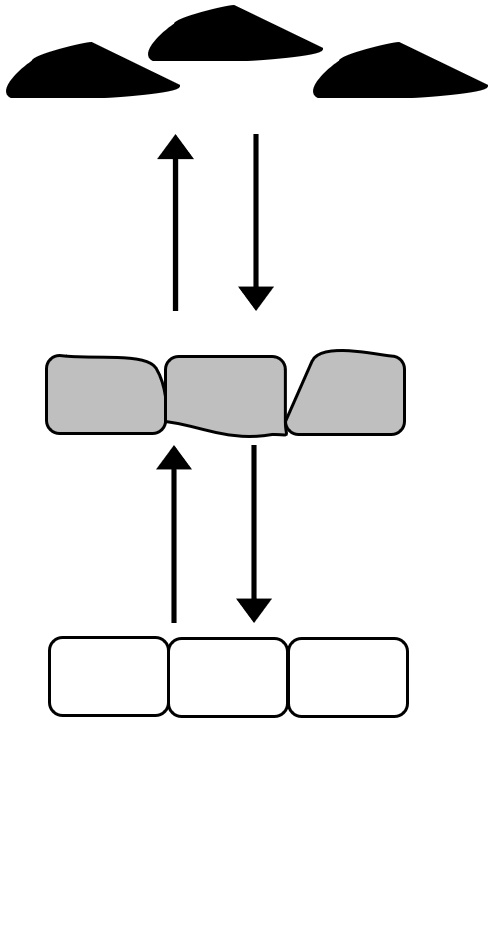}
\caption{ }
\label{fig:16b}
\end{subfigure}
\caption{(a) Time-evolution exclusively through monostable regions. Every point on the path is monostable.(b) Image of monostable cells. Pro-metastatic cells in black,  Anti-metastatic cells in white and intermediate monostable cells in continuous transition region in grey.}
\label{fig:16}
\end{figure}

In Figure \ref{fig:16} we have superimposed potential paths of $\alpha(t)$ in the space of parameters $\rho$ and $k$ from Figure \ref{fig:1}.

The time-evolution of the parameters $\alpha(t) =( \rho, k )$ shows possible transitions between stable cell types. Here, the green line indicates how values of $\alpha(t)$ begin in the monostable anti-metastatic region (A), traverses an ambiguous monostable region and arrives into the monostable pro-metastatic region (P). These transitions we will call APT, or PAT depending on the direction. The unique stable state of the differential equations $x(t)$ evolves together with $\alpha(t)$ but the possible cellular type ultimately transitions between pro-metastatic and anti-metastatic states. Therefore this image represents a continuous transition depending on which way the path is traversed. It is important to note that both transitions traverse the ambiguous region to the left of the bistable region. Cells in this ambiguous region are neither pro- nor anti-metastatic. They are in an intermediary, biologically ambiguous cell state where the pro-and antimetastatic states become indistinguishable. Such ambiguous parameter regions must also appear in all other analyses describing bistable or multistable systems, but they have not been mathematically characterized from the perspective of metastasis. Analogies may exist with the physics of phase transitions, e.g., liquid and gas phases becoming indistinguishable beyond the critical point of water.

\begin{figure}[H]
\begin{subfigure}{0.4\textwidth}
\centering
\includegraphics[scale=.5]{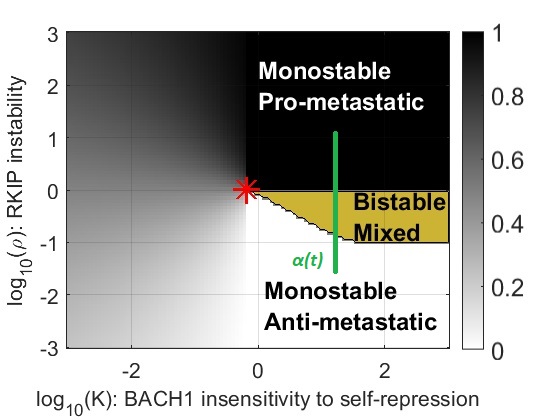} 
\caption{ }
\label{fig:17a}
\end{subfigure}
\begin{subfigure}{0.4\textwidth}
\centering
\includegraphics[scale=.2]{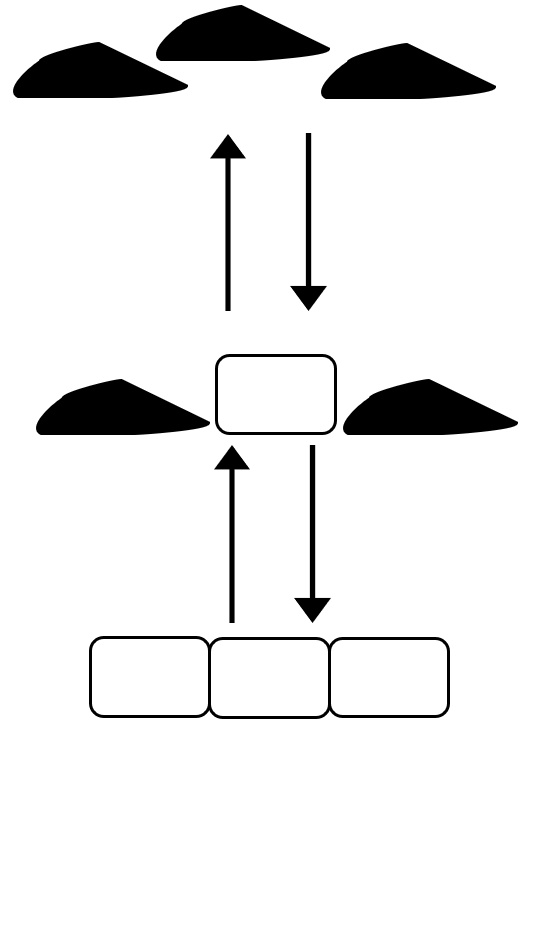}
\caption{ }
\label{fig:17b}
\end{subfigure}
\caption{(a) Transition between monostable regions traversing bistable mixed region. (b) Image of Discontinuous transition. Monostable Pro-metastatic cells in black, Monostable Anti-metastatic cells in white and cells in mixed Bistable region in both white and black.}
\label{fig:17}
\end{figure}

In Figure \ref{fig:17} the path of $\alpha(t)$ crosses the boundary between the monostable and bistable regions. Similar transitions have been extensively described in the EMT/MET literature. At first we consider the path as it goes up transitioning from anti-metastatic to pro-metastatic (APT). As the path crosses the boundary at $t_0$, the edge of the bistable mixed region, the stable state $x(t)$ continues on while a new equilibrium point is created which we denote $x'(t_0)$. Over time the point $x'(t)$ splits into a new stable point $x'_1(t)$ and a saddle $x'_2(t)$. In the bistable region there are three equilibria $x(t)$, $x'_1(t)$ and $x'_2(t)$. If we cross the higher boundary of the bistable region a stable point and a saddle collide and annihilate each other that is, $x(t)$ and $x'_2(t)$ collide.

This would be a discontinuous transition in the direction A to P, since at every point along the path where the A and P states co-exist, they correspond to separate sets of variables $x_A$ and $x_P$, which are clearly distinguishable. There is no ambiguity at any point about an individual cell being in one state or another. But now as we run this path backwards to produce a PAT we see that the state of the cell corresponds to $x'_1(t)$ as $x(t)$ tends to $x'_1(t)$ in the bistable region. Thus traversing the path in one direction and then the other exhibits hysteresis. Hysteresis is another hallmark of discontinuity. Unlike continuous transitions, discontinuous EMT/MET has been extensively investigated for many regulatory networks (See also Figure \ref{fig:4}). 

Here is a picture in 1-dimensions of what is happening.

\begin{figure}[h]
\centering
\includegraphics[scale=.7]{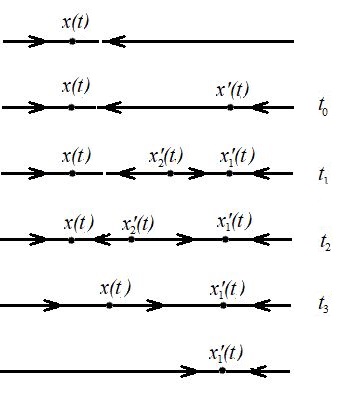}
\caption{1-dimensional time phases of stable states}
\label{fig:18}
\end{figure}

Now if we imagine that metastasis will require an APT transition followed by a PAT transition we see that the PAT transition may require that the parameter cross the whole bistable region in reverse. The region near or left of the star in Figure \ref{fig:1} would seem to provide the most fertile region for such a transition in both directions. This applies to both continuous and discrete transitions, to the left and right of the red star, which indicates the cusp point. Two questions are immediate:

1. What determines the path $\alpha(t)$ of continuous or discontinuous transition that the cell will trace out in the parameter space?  

2. Will the paths of the cells stay away from or pass close to the cusp point? What is the biological significance of this alternative? Will paths that pass close to the cusp be more likely to undergo an APT transition, followed by an PAT transition, and thus be more likely to establish distant tumors? 

We addressed these questions in connection with metastatic cell state transitions above. We also proposed the possibility of seeing the same behavior in 2 or 3-gene circuits and in cell division dynamics that mimic a ``Toggle" network (See supplementary materials). A general and in depth perspective on the mathematics behind finding a cusp bifurcation is introduced in the appendix which we hope could be applied to other biological networks. It is possible  even more biological systems exhibit cusp and pitchfork bifurcations where a simple one-parameter binary transition may not sufficiently describe the biological phenomenon and we think that it is possible to show mathematically that these systems can undergo two-parameter bifurcations. If this is the case then we believe it is important to revisit existing studies of gene networks and apply these findings from two-parameter bifurcation theory to system models and predictive techniques.

We note that there is another approach to the use of bifurcation theory in biology. This is the Catastrophe theory of Rene Thom \cite{ref4thom2}. The developed theory concerns the zeros of gradient vector fields and their bifurcations. There is a cusp bifurcation which is very much the same in its geometric features. However, one has to be a little careful here since the bifurcation theory of differential equations and gradient differential equations have some subtle differences. For example the general cusp bifurcation has the possibility of exhibiting a periodic solution which the gradient system does not. The survey paper by Rand et al \cite{rand} has recent updates to this theory. Given a differential equation satisfying certain properties, there is a gradient system which shares the asymptotic behavior of the original system. Now Rand adds the unstable manifold geometry behavior to the analysis. A drawback of this theory may be that the gradient system is not immediately at hand, whereas we work with the equations directly.

Further work as it relates to metastatic breast cancer would be to incorporate these models to existing predictive processes that assess metastatic potential. Furthermore, it would be valuable to validate similar cusp behaviors are found across other metastatic cancers more generally. Further interesting mathematical studies would be to solve the equations in Lee et al., and other existing metastasis models, symbolically to make sure we have all possible solutions. It would also be valuable to rigorously prove that the rank of the derivative is always 4 in such systems to validate our hypothesis. A forthcoming analysis of these initial results could be made to generalize the observation of cusp bifurcations in other bistable or multistable biological networks and extend two-parameter bifurcation theory theory in a variety of biological phenomenon. 

\section{Acknowledgments}

This research was supported by the Rich Summer Internship program and City College. Michael Shub's research was partially supported by the Smale Institute. GB acknowledges support by the National Institutes of Health, NIGMS MIRA Program (R35GM122561) and by the Laufer Center for Physical and Quantitative Biology.

We thank Indika Rajapakse for helpful discussion as well as his software recommendation, and we thank him for pointing out some of the literature concerning cell division as well as open problems to us. We also thank Chris York's assistance in producing a final cusp fold image in MATLAB.

\newpage

\newpage

\section{Appendix A}
Here we present some elementary examples of biological systems that exhibit cusp bifurcations. We include an extensive analysis of the mathematics involved in solving such systems for those who are interested.

\subsection{Cusp Bifurcation in biological networks}

\subsubsection{Gene circuits}

The cusp point is relevant to many areas of biology besides metastasis. It should be found for gene networks described earlier as bistable or multistable. This applies to the toggle switch, one of the foundational gene circuits in synthetic biology. Thus, we considered the 2-gene and a 3-gene networks from Appendix 1A and 1C of Smale and Rajapakse \cite{rajapakse2016mathematics}, as other examples of a cusp bifurcation in biological systems. Smale and Rajapakse showed that pitchfork bifurcations exist for these networks. Below we give an analysis of the 2-gene network. 

\begin{figure}[h]
\begin{subfigure}{0.4\textwidth}
\includegraphics[scale=.4]{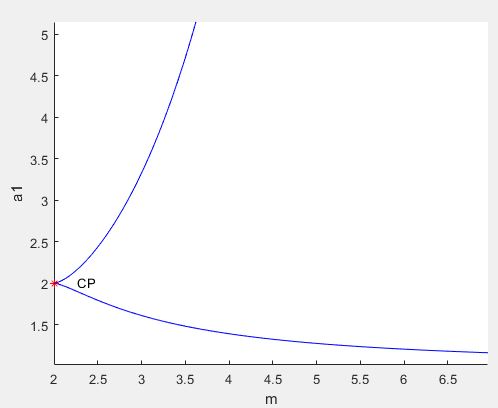}
\caption{Plot varying $\alpha_1,m$ where $\alpha_2=2$}
\label{fig:2-gene pitchfork m,x}
\end{subfigure}
\begin{subfigure}{0.4\textwidth}
\includegraphics[scale=0.4]{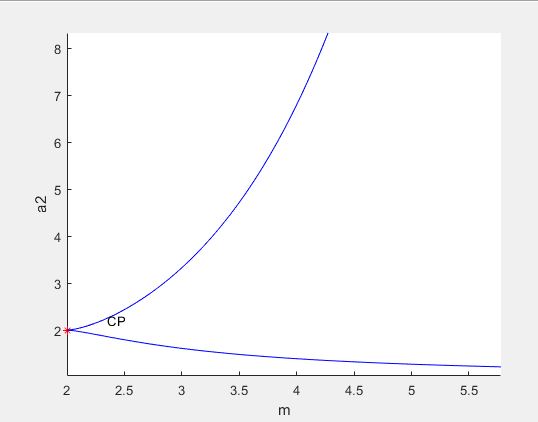}
\caption{Plot varying $\alpha_2,m$ where $\alpha_1=2$}
\label{fig:2-gene pitchfork m,y}
\end{subfigure}
\caption{MATCONT 2-gene cusp results}
\label{fig:MATCONT}
\end{figure}

\begin{figure}
    \centering
    \begin{subfigure}{0.4\textwidth}
        \includegraphics[scale=.4]{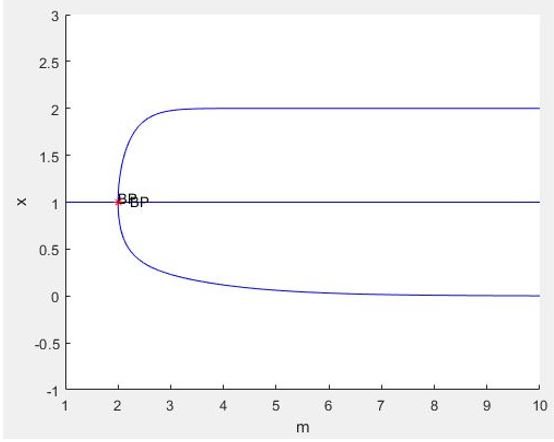}
        \caption{Plot of $x,m$}
        \label{fig:2gene PF1}
    \end{subfigure}
    \begin{subfigure}{0.4\textwidth}
        \includegraphics[scale=.4]{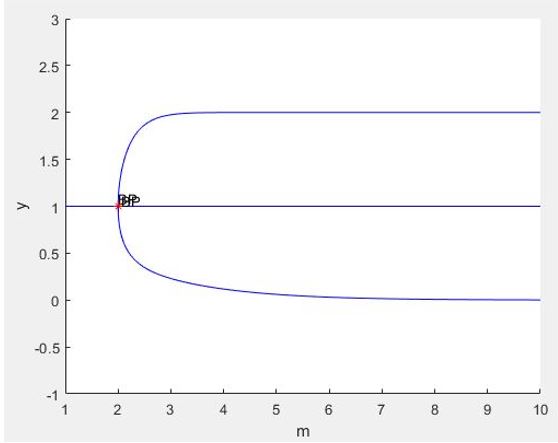}
        \caption{Plot of $y,m$}
        \label{fig:2gene PF2}
    \end{subfigure}
    \caption{MATCONT pitchfork bifurcation}
    \label{fig:2gene PF}
\end{figure}

We used MATCONT to show that a cusp point occurs for the system 
\begin{align*}
\dot{x}&=\frac{\alpha_1}{1+y^m}-x\\
\dot{y}&=\frac{\alpha_2}{1+x^k}-y    
\end{align*}
In Figure \ref{fig:2-gene pitchfork m,x} we set $\alpha_2 = 2$ and $m=k >0$. As we varied $\alpha_1$ and $m$ we found a cusp exists at $(x,y,\alpha_{1},m)=(1.00,1.00,2.00,1.99)$. In Figure \ref{fig:2-gene pitchfork m,y} we set $\alpha_1 = 2$, $m=k >0$ and varied $\alpha_1$ and $m$. A cusp point exists at $(x,y,\alpha_{2},m)=(1.00,1.00,2.00,1.99)$. This aligns with Smale and Rajapakse's results \cite{rajapakse2016mathematics}. They set $\alpha_1=\alpha_2 =2$, $m=k >0$ and proved that a pitchfork bifurcation occurs at the point $m =2$. They note that for all $0\leq m <2$ there is only one solution to the one parameter system which is $(x,y) = (1,1)$. It can be shown that for values past the cusp, or when $m>2$, there are three solutions and we have the same pitchfork scenario as we described  in the paper. We plot the pitchfork using MATCONT in Figure \ref{fig:2gene PF}.

\begin{figure}[H]
\centering
\begin{subfigure}{0.4\textwidth}
\includegraphics[scale=.4]{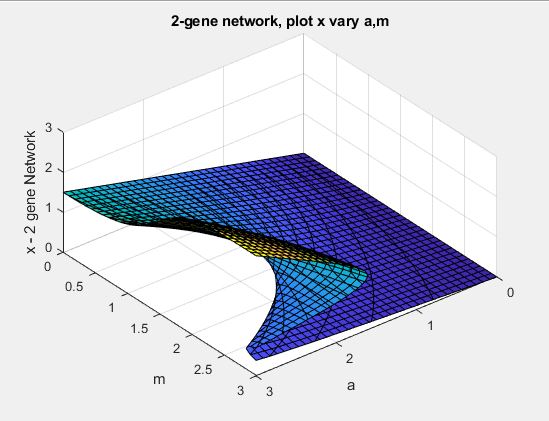}
\caption{2-gene x, $\alpha_1,m$ plot}
\label{fig:fimplicit3 plot of x fold}
\end{subfigure}
\begin{subfigure}{0.4\textwidth}
\includegraphics[scale=0.4]{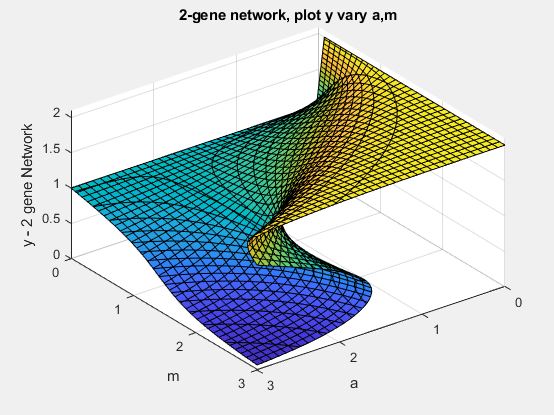}
\caption{2-gene y, $\alpha_2,m$ plot}
\label{fig:fimplicit3 plot of y fold}
\end{subfigure}
\caption{Matlab fimplicit3 plots}
\label{fig:fimplicit3 plots of x and y}
\end{figure}

In Figure \ref{fig:fimplicit3 plots of x and y} we used the same fimiplicit3 plot in MATLAB as we did in the paper. We plot the solution surface of $x$ and $y$ separately and vary the two relevant parameters  $(\alpha_1,m)$ and  $(\alpha_2,m)$ respectively. The surfaces both are folds which we would expect to see for a cusp bifurcation. Figure \ref{fig:MATCONT} shows the projection of the fold on the $(\alpha_{1},m)$ and $(\alpha_{2},m)$ plane. As for the metastasis-regulatory network, the cusp point separates the parameter ranges for continuous versus discontinuous transitions.

\subsubsection{Cell Division}

Another important network for which our analysis applies drives the cell cycle. Rajapakse and Smale \cite{rspfb} suggest that the pitchfork bifurcation may explain some of the properties of cellular division. The decision of cells between cycling and senescence is driven by a bistable network, the core of which is a variant of the toggle switch involving CDK2 and Rb (Figure 2.A in \cite{spenceretal}). Considering the above analysis for the toggle switch, the cell cycle network might also be modelled as a Toggle, with a pitchfork or cusp bifurcation. Thus, we use the 2-gene Toggle network from Smale and Rajapakse \cite{rajapakse2016mathematics} to demonstrate what could be happening during cell division and how it relates to a cusp bifurcation. We use the toggle switch to illustrate some biologically relevant features of the dynamics and note that the bifurcations of the cell cycle network will occur at different, currently unknown parameter values.

Since the toggle switch has two steady states, one of which corresponds to entering the cell cycle, we need to assign a biological meaning to the other steady state. We propose that the other steady state corresponds to differentiation, which is known to be antagonistic with cell cycling \cite{PMID:24979803} \cite{PMID:19666818}.

The dynamics of a differential equation with fixed parameters after the pitchfork or cusp bifurcations are:

\begin{enumerate}
    \item There are 3 equilibria, two sinks and a saddle.
    \item The forward orbit of any point tends to one of the equilibria.
    \item The orbits which tend to the saddle, called the stable manifold of the saddle, form an (n-1) dimensional disc which separates the two basins of the stable equilibria.
\end{enumerate}

\begin{figure}[H]
\centering
\includegraphics[scale=.6]{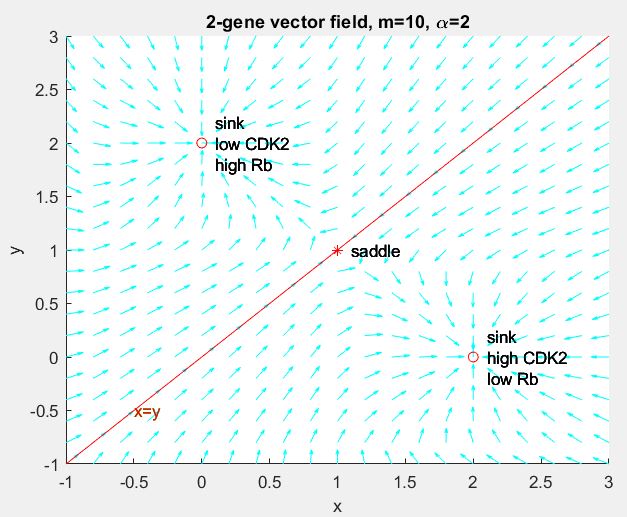}
\caption{Vector plot of 2 gene circuit system to show 3 equilibria \cite{rajapakse2016mathematics}}
\label{fig:saddlesinks}
\end{figure}

In Figure \ref{fig:saddlesinks} we produce a vector plot using the 2-gene Toggle network from Smale and Rajapakse \cite{rajapakse2016mathematics}. We see all three behaviors listed in this image, where the upper sink could be low CDK2/high Rb, the lower sink is high CDK2/low Rb and the $x=y$ line divides paths a cell may take as they differentiate if they initiate above the line or proliferate if they begin on a path below the line.

Almost all orbits tend to one of the stable equilibria. An orbit starting near the red line will have very slow dynamics. Paths close to the unstable equilibria and far from the stable equilibria will take a long time to arrive near a stable equilibrium. In the case of cell division one stable equilibrium corresponds to the cell being proliferative, and the other to cell differentiation. Points close to the stable manifold will have fast dynamics as they tend to the sink. However, points close to the unstable equilibria and far from the stable equilibria will define cells that may appear quiescent for a long time, although ultimately they will commit to cycling. Cells distributed along the green arrows could have drastically long waiting times before entering the cell cycle, as observed in the experiments.

\begin{figure}[H]
\centering
\begin{subfigure}{0.4\textwidth}
\includegraphics[scale=.4]{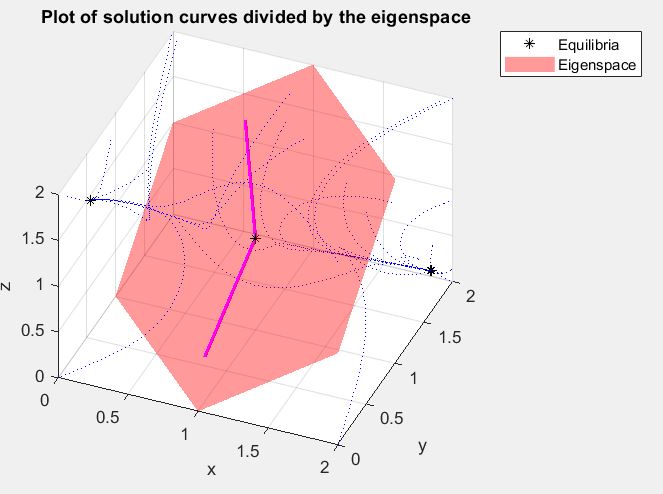}
\caption{ }
\label{fig:eig1}
\end{subfigure}
\begin{subfigure}{0.4\textwidth}
\includegraphics[scale=0.3]{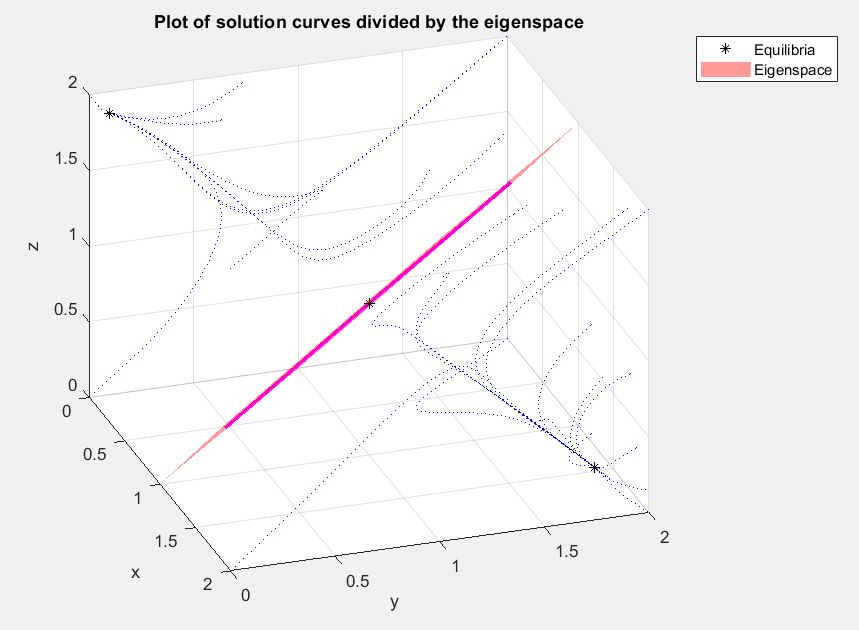}
\caption{ }
\label{fig:eig}
\end{subfigure}
\caption{Matlab plot of eigenspace and solution curves in blue. Lines in magenta are the eigenvectors of the saddle point which determine the eigenspace.}
\label{fig:eigs}
\end{figure}

We see the same behavior in 3 dimensions. The 3 dimensional system is defined by Smale and Rajapakse as \cite{rajapakse2016mathematics} follows:
$$\dot{x}=\frac{\alpha}{1+z^m}-x$$
$$\dot{y}=\frac{\alpha x^m}{1+x^m}-y$$
$$\dot{z}=\frac{\alpha}{1+y^m}-z$$
To find the behavior in the bistable region we set $\alpha = 2$ and $m=3$. In Figure \ref{fig:eigs} we show the 3 equilibria as black stars, and the forward orbits are the blue dotted lines each tending towards one of the equilibria. The eigenspace of the saddle is a plane that divides two regions in three dimensions, separating the two basins of stable equilibria. This plane is an approximation of the stable manifold of the saddle. We use \texttt{ode45} in MATLAB to solve the system starting from various points in the space. Most points stay on the same side of the manifold as where they started. 

We suggest that the same behavior is seen in the system involving CDK2, Rb and Ef2 (Figure 2.A in \cite{spenceretal}). In our model we may interpret $x,y,z$ as EF2, CDK2, Rb respectively. In this space as $z$ increases $x,y$ decrease and solutions tend to the stable sink in the upper left hand corner of Figure \ref{fig:eig}, which corresponds to differentiation. Conversely, as $z$ decreases, $x,y$ increase and solutions tend to the sink in the lower right hand corner of Figure \ref{fig:eig}, which corresponds to cell cycle entry. This reflects a switch behavior in cell division with 3 variables. 

There is a symmetry in the 2-gene model when $m> 2, \alpha_1=\alpha_2$. In two dimensions the line $x=y$ is the stable manifold that separates the two cell behaviors. Any initial point with $x>y$ or $x<y$ will converge to the same stable equilibirum. Similarly when considering the 3 dimensional model the stable manifold of the saddle separates the space such that all orbits tend to the sink of the same side. While our model is the linear approximation of the stable manifold it is clear that for values of $x$ that are large enough, all orbits tend to the stable state which corresponds to high CDK2 concentrations. This is the observed situation with CDK2 and proliferation \cite{spenceretal}. In the examples the line in 2-dimension and the disc or eigenspace in 3-dimension approximate the true manifold and are based off simple gene networks, however, it could be interesting to develop a more accurate model of cell division to combine the findings in the two papers referenced \cite{rajapakse2016mathematics,spenceretal}. Furthermore, one could demonstrate that a pitchfork or cusp bifurcation exists in cell division as well.

\section{Appendix B: Cusp Bifurcation Methods}

So far our analysis has been based on the rather fine numerical work of Lee et. al.\cite{metarticle} represented in Figure 1 of the paper. But we may give a more analytical treatment of the figures which in this case and much more generally establish the existence of the cusp point and give a sharper description of the geometry of the division into monostable and multistable regions near it. The discussion of how the stable states change along paths in the parameter space will be the same in this greater generality. The subject matter is called bifurcation theory. We are given smooth differential equations $\dot{x}=V(x,\alpha)$ as above where $x \in \mathbb{R}^n$ and $\alpha$ is a parameter in $\mathbb{R}^j$ and $V:\mathbb{R}^n \times \mathbb{R}^j \to \mathbb{R}^n$. In our case $\alpha$ is a real parameter so $j=1$.

Bifurcation theory studies how the behavior of the solution of $V$ changes as $\alpha$ changes. This is a very well developed subject \cite{guckenheimer, kuznetsov,govaerts}. The two stable bifurcations which will interest us are the saddle-node and cusp bifurcations. We will use the referenced numerical methods to find them. \cite{saddlewebsite, cuspwebsite} 
The equation 1) $V(x,\alpha)=0$ defines the equilibria. Recall that in general $$V(x,\alpha): \mathbb{R}^n \times \mathbb{R} \to \mathbb{R}^n$$ and we assume that when $V(x,\alpha)=0$ 

equation 2) $D V(x,\alpha)$ goes from $\mathbb{R}^{n+1} \to \mathbb{R}^n$ and the rank of the derivative is $n$. 

So the set of zeros is locally a curve near $(x,\alpha)$ by the implicit function theorem. 

We assume one of the eigenvalues of the derivative $D_x V(x,\alpha)$ is $0$, so we add 

equation 3) $Det[D_x V(x,\alpha)]=0$ 

and now we assume that $$(V(x,\alpha),Det[D_x V(x,\alpha)]) : \mathbb{R}^n \times \mathbb{R} \to \mathbb{R}^n \times \mathbb{R}$$ has rank $n+1$. 

So $x,\alpha$ is an isolated solution of equations 2) and 3), and the curve of zeros is tangent to the $\alpha=$constant plane. A saddle of index $k$ and one of index $k-1$ collide and annihilate or appear if the parameter runs in the opposite direction.

Here is an illustration of the saddle node bifurcation from Scholarpedia in one dimension and one parameter.

\begin{figure}[H]
\centering
\includegraphics[scale=.7]{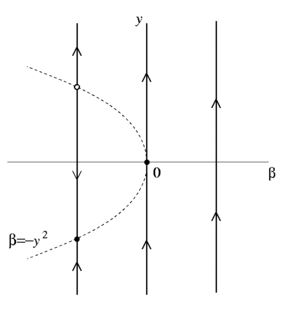}
\caption{Saddle-node bifurcation in the one-dimensional system \cite{saddlewebsite}}
\label{fig:19}
\end{figure}

If $n=2$ or higher, here is the dynamical picture in the plane. In general the downward arrow represents an $(n-1)$ dimensional plane of the non-zero eigenvalues.

\begin{figure}[H]
\centering
\includegraphics[scale=.5]{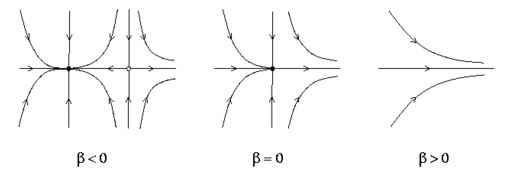}
\caption{Saddle-node bifurcation on the plane in the system \cite{saddlewebsite}}
\label{fig:20}
\end{figure}

Now if we have two parameters $\rho$ and $k$ $$V(x,\rho,k): \mathbb{R}^n \times \mathbb{R}^2 \to \mathbb{R}^n$$ Then 

1$'$)$V(x,\rho,k)=0$ 

defines the set of equilibria and we assume 

2$'$)the rank of $D(V(x,\rho,k))=n$ 

so the set of equilibria near $(x,\rho,k)$ is a 2-dimensional surface. On the surface we have those equilibria with a zero eigenvalue where the number of equilibria might be changing, these are described by adding the equation 

3$'$) $Det[D_x V(x,\rho,k)]=0$ 

and we assume that the map $(x,\rho,k)\to (V(x,\rho,k), Det[D_x V(x,\rho,k)])$ has rank $n+1$ where the two equations are satisfied. Hence the solution is a curve near $(x,\rho,k)$.

When we project the curve on the parameter plane, $(\rho,k)$ there is either A) no singularity of the projection, i.e. no tangent vector to the curve is vertical or B) there is a singularity of the projection, that is a tangent vector to the curve which is vertical.

\begin{figure}[h]
\centering
\includegraphics[scale=.7]{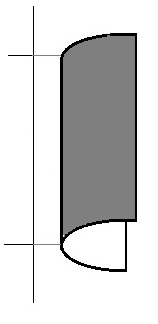}
\caption{Fold projected on parameter plane}
\label{fig:21}
\end{figure}

In case A) there are two possibilities:

Aa) 0 is a simple eigenvalue of $D_x(V(x, \rho,k)$ along the curve. In this case the projected curve divides the plane near $(\rho,k)$. On one side there is no equilibrium on the other side, there are two of index $j$ and $j-1$. So in a region of interest to us, a saddle and a sink. If a curve in the parameter plane crosses the projected curve we see a saddle-node bifurcation.

This is called a fold bifurcation.

Ab)There is a point along the curve where the rank of $D_x V(x, \rho,k)$ is $n-1$ but a second eigenvalue is 0. Now when a curve crosses the projected curve we may create a saddle of index $j$ and another equilibria of index $j-1$ or $j+1$ depending on where the crossing takes place and for some values of the parameter there are limit cycles. We don’t see this behavior in the data of Lee et al. and we will not discuss it. This bifurcation is called a Bogdanov-Takens bifurcation.

There are generically five types of bifurcations of equilibria depending on two parameters (See Kuznetsov). All might be interesting for biology.

Case B) is the one that interests us. It is the cusp bifurcation. There is a singularity of the projection, that is a tangent vector to the curve which is vertical. Recall that the tangent to the curve is the null space of the derivative of $(V(x,\rho,k),Det[D_x V(x,\rho,k)])$. Now for a vertical vector $(v,0,0)$ to be in the null space it is necessary and sufficient that

4$'$) $D_x(V(x,\rho,k)(v)=0$ and

5$'$) $\nabla_x Det[D_x V(x,\rho,k)]\bullet (v) =0$

The references above give methods for determining that these equations have a solution. When equations 1$'$) and 3$'$) are satisfied $D_xV(x,\rho,k)$ has a kernel as $Det[D_xV(x,\rho,k)]=0$ so we would like to express the vector $v$ in terms of $(x,\rho,k)$. Assuming that $D_xV(x,\rho,k)$ has rank $n-1$ when $V(x,\rho,k)=0$ then at least one of the $(n-1)$ by $(n-1)$ minors has non-vanishing determinant and the adjugate matrix is not zero. By Cramer's Rule any non-zero column of the adjugate matrix is in the kernel of $D_x V(x,\rho,k)$.

So a cusp point is defined as a non-degenerate solution of the $n+2$ by $n+2$ system of equations
\begin{equation}
    \begin{split}
        V(x,\rho,k)&=0 \\
        Det[D_x V(x,\rho,k)]&=0\\
        \nabla_x Det[D_x V(x,\rho,k)]\bullet (v)&=0
\end{split}
\end{equation}

Where $v$ is a non-zero column of the adjugate matrix of $D_x V(x,\rho,k)$.

Here are some pictures of the cusp bifurcation when $n=1$.

\begin{figure}[H]
\centering
\includegraphics[scale=.6]{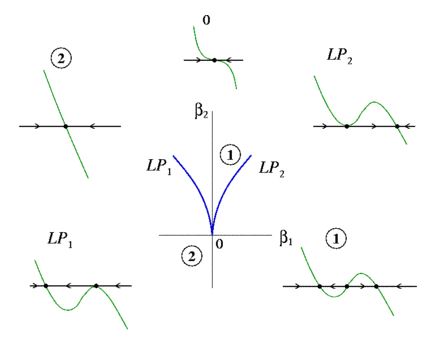}
\caption{Cusp bifurcation in the one-dimensional system \cite{cuspwebsite}}
\label{fig:22}
\end{figure}

The surface below is the surface of equilibria. Outside of the cusp region there is one equilibrium and inside 3.
\begin{figure}[H]
\centering
\includegraphics[scale=.4]{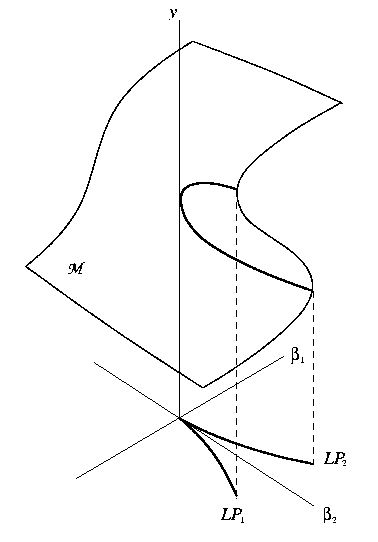}
\caption{Equilibrium manifold near a cusp bifurcation \cite{cuspwebsite}}
\label{fig:23}
\end{figure}

As a path crosses the cusp curve away from the cusp point itself the equation has a saddle-node bifurcation. The picture of cubics illustrates again what is happening in 1-dimension. The two curves of the cusp are tangent at the cusp point. So the cusp point is a place where the hysteresis effect may in some sense be briefest. See questions 1 and 2 above.

A strategy for the analysis of equilibria might be to look for non - degenerate solutions of $V(x,\alpha)=0$ and $Det[D_x V(x,\alpha)]=0$ for saddle node bifurcations and of the system of equations (2) where $D_x V(x,\alpha)(v)=0$, $Det[D_x V(x,\alpha)]\bullet (v)=0$ and $v$ is a column of the adjugate matrix of $D_x V(X,\alpha)$ for cusp bifurcations. The region in Figure 1 looks like there might be a cusp bifurcation which puts a sharper definition of the regions and supports the cell transitions in Figure 2 of the paper.

\section{Appendix C: Matlab Code}

Figure 1
\begin{verbatim}
%function zB = funB(t,B)
function zB = funB(B)
%no shBACH1 here

global s S c K m b r rho a

zB = s+(S-s)*K^b/(K^b+B^b) - B*( 1+ (a/(m^r*rho^r*(1+B)^r+1)) * (c/(1+c*B)) );

function [Ld,Lt] = LOSSderiv(B)

global s S c K m b r rho a

% loss term
Lt = B.*(1+ (a./(m^r*rho^r*(1+B).^r+1)).*(c./(1+c*B)));

fderiv_num1=(m^r*rho^r*(1+B).^r+1).^2.*(1+c*B).^2;
fderiv_num2=a*c*(m^r*rho^r*(1+B).^r+1).*(1+c*B);
fderiv_num3=a*c^2*B.*(m^r*rho^r*(1+B).^r+1);
fderiv_num4=a*c*m^r*rho^r*r*B.*(1+B).^(r-1).*(1+c*B);
fderiv_den=(m^r*rho^r*(1+B).^r+1).^2.*(1+c*B).^2;

% fd1=1+a*c./((m^r*rho^r*(1+B).^r+1).*(1+c*B));
% fd2=B.*(a*c^2*(m^r*rho^r*(1+B).^r+1))./((m^r*rho^r*(1+B).^r+1).^2.*(1+c*B).^2);
% fd3=B.*(a*c*m^r*rho^r*r*(1+B).^(r-1).*(1+c*B))./((m^r*rho^r*(1+B).^r+1).^2.*(1+c*B).^2);


% derivative of loss term
Ld=(fderiv_num1+fderiv_num2-fderiv_num3-fderiv_num4)./fderiv_den;
%Ld=fd1-fd2-fd3;

clear all; close all;

global s S c K m b r rho a alpha

L=3;
paramrange=logspace(-L,L,100);

B=logspace(-5,3,100);

colorsetR=[0.5 0 0; 1 0 0; 1 .5 .8];
colorsetB=[0 0 0.5; 0 0 1; 0 1 1];

clear isbist;clear lowBACH1;
s=0.02;
S=20;
c=200;
m=2;
b=3;
r=5;
a=1000;
% scan rho: RKIP degradation rate
% scan alpha: shBACH1 effect
Krange=paramrange;
rhorange=paramrange;
ctr1=1;
%figure;
for rho=rhorange
    ctr2=1;
    for K=Krange
        lhs=B.*(1+ (a./(m^r*rho^r*(1+B).^r+1)).*(c./(1+c*B)));
        rhs=s+(S-s)*K^b./(K^b+B.^b);
        %subplot(7,7,(ctr1-1)*7+ctr2);plot(B,lhs,'LineWidth',2);hold on;plot(B,rhs,'r','LineWidth',2);set(gca,'XScale','Log','YScale','Log');

        [Ld,Lt]=LOSSderiv(B);   % get the derivative of the loss term, and the loss term

        seekdownslope=1-sign(Ld);   %TRUE is loss term's derivative is negative or 0
        midP=mean(B(find(seekdownslope)));
        %[a K any(seekdownslope) 3]
        
        %chance for bistable region next
        if(any(seekdownslope) && Ld(floor(mean(find(seekdownslope))))*Ld(end)<0 && Ld(floor(mean(find(seekdownslope))))*Ld(1)<0)
            
            zD1=fzero('LOSSderiv',[B(1) midP]);
            zD2=fzero('LOSSderiv',[midP B(end)]);

            [Ld1,Lt1]=LOSSderiv(zD1);
            [Ld2,Lt2]=LOSSderiv(zD2);
            Gt1=(s+(S-s)*K^b./(K^b+zD1.^b));    % gain term 1
            Gt2=(s+(S-s)*K^b./(K^b+zD2.^b));    % gain term 2
            isbist(ctr1,ctr2)=(Lt1-Gt1>0)*(Gt2-Lt2>0)*(zD2-zD1>0);
            lowBACH1(ctr1,ctr2)=(Gt2-Lt2<0);
            %[a K (Lt1-Gt1) (Gt2-Lt2)]
            %sBACH1a(ctr1,ctr2)=Gt1-Lt1;  % gain term - loss term = dB/dt
            %sBACH1b(ctr1,ctr2)=Gt2-Lt2;  % gain term - loss term = dB/dt
        %monostability is certain
        else
            isbist(ctr1,ctr2)=0;
            zD=fzero('funB',[0 B(end)*100]);
            lowBACH1(ctr1,ctr2)=(zD<0.2);
            %sBACH1(ctr1,ctr2)=zD<0.2;
        end;
        ctr2=ctr2+1;
    end;
    ctr1=ctr1+1;
end;

xvec=log10(rhorange);yvec=log10(Krange(1:47));
[xx,yy]=ndgrid(xvec,yvec);
% figure;mesh(xvec,yvec,atan2(yy,xx));
atval=atan2(yy,xx);
atval(48,length(Krange))=0;

% figure;imagesc(log10(rhorange),log10(Krange),atval/min(min(atval)));colorbar;
% figure;imagesc(log10(rhorange),log10(Krange),isbist);colorbar;

B = bwboundaries(isbist);
%mask=isbist>0.5;

bistsurf=isbist*0.5+lowBACH1;
%bistsurf=lowBACH1;
%bistsurf(:,1:47)=(atval(:,1:47)/min(min(atval))).^0.75;
bistsurf(:,1:47)=(atval(:,1:47)/min(min(atval)));

figure;
I=imagesc(log10(rhorange),log10(Krange),bistsurf);
patch(log10(Krange(B{:}(:,2))),log10(rhorange(B{:}(:,1))),[0.8 0.7 0.2])
%imagesc(log10(rhorange),log10(Krange),sBACH1);colorbar;
hold on;cb=colorbar;
%patch(log10(rhorange(isbist)),log10(Krange(isbist)),'g');

% plot(log10(10),log10(.05),'o','Color',colorsetB(1,:),'LineWidth',2,'MarkerSize',20);
% plot(log10(10),log10(.5),'o','Color',colorsetB(2,:),'LineWidth',2,'MarkerSize',20);
% plot(log10(10),log10(5),'o','Color',colorsetB(3,:),'LineWidth',2,'MarkerSize',20);
% 
% plot(log10(10),log10(.5),'^','Color',colorsetR(3,:),'LineWidth',2,'MarkerSize',16);
% plot(log10(.1),log10(.5),'^','Color',colorsetR(1,:),'LineWidth',2,'MarkerSize',16);
% plot(log10(.1),log10(5),'^','Color',colorsetB(3,:),'LineWidth',2,'MarkerSize',16);

plot(log10((Krange(47)+Krange(48))/2),log10((rhorange(50)+rhorange(51))/2),'*','Color','r','LineWidth',2,'MarkerSize',20);
[(Krange(47)+Krange(48))/2 (rhorange(50)+rhorange(51))/2]

set(gca,'FontSize',16,'YDir','normal');
xlabel('log_{10}(K): BACH1 insensitivity to self-repression','FontSize',16);
ylabel('log_{10}(\rho): RKIP instability','FontSize',16);
colormap(gray);grid on;caxis([0 1]);
text(0,2,{'Monostable','Pro-metastatic'},'Color',[1 1 1],'FontSize',18,'FontWeight','Bold');
text(0.1,-2.2,{'Monostable','Anti-metastatic'},'Color',[0 0 0],'FontSize',18,'FontWeight','Bold');
text(1.5,-.5,{'Bistable','Mixed'},'Color',[0 0 0],'FontSize',18,'FontWeight','Bold');
set(cb,'TickLabels',flipud(get(cb,'TickLabels')))

\end{verbatim}

Figure 3
\begin{verbatim}
function [lastDW,lastN,N,Wx,DWd,DWw]=NewtonsMethodfcn(eqns,x0,vars,count)

NTable=[];
WxTable=[];
vars0=x0;
W=double(subs(eqns,vars,vars0)); 
DW=jacobian(eqns,vars); 
DWsub=double(subs(DW,vars,vars0));

if rank(DW)==size(DW,2)
    DWdagger=inv(DWsub);
else
    DWdagger=DWsub'*inv(DWsub*DWsub');
end
DWW=DWdagger*W;
Nprime=vars0'-DWW; 
WNprime=subs(eqns,vars,Nprime'); 
NTable=[vars0',Nprime];
WxTable=[W,WNprime];
for j=1:count
    W=double(WNprime);
    DWsub=double(subs(DW,vars,Nprime')); 
    if rank(DW)==size(DW)
        DWdagger=inv(DWsub);
    else
        DWdagger=DWsub'*inv(DWsub*DWsub');
    end
    DWW=DWdagger*W;
    Nprime=Nprime-DWW;
    NTable=[NTable,Nprime];
    WNprime=double(subs(eqns,vars,Nprime'));
    WxTable=[WxTable,WNprime];    
end
lastDW=DWsub;
lastN=Nprime;
N = double(NTable);
Wx = double(WxTable);
DWd=DWdagger;
DWw=DWW;

%script to generate table using Newtons Method function
s=.02;
S=20;
c=200;
m=2;
b=3;
r=5;
a=1000;

syms R L B rho K

assume([R L B], 'Real')
assume([R L B], 'Positive')

eqn1=(1/(1+B))-rho*R;
eqn2=((a*R^r)/(m^r+R^r))-L-c*L*B;
eqn3=s+(((S-s)*K^b)/(K^b+B^b))-B-c*L*B;

%a11*(a22*a33 - a23*a32) + a13*a21*a32 from Jacobian matrix
eqn4=(-rho)*((-B*c-1)*(-L*c-(B^(b-1)*K^b*b*(S-s))/(B^b+K^b)^2-1)
    -(-L*c)*(-B*c))+(-1/(B+1)^2)*((R^(r-1)*a*r)/(R^r+m^r)-(R^r*R^(r-1)
    *a*r)/(R^r+m^r)^2)*(-B*c);
eqns=[eqn1;eqn2;eqn3;eqn4];
vars=[R,L,B,rho,K];

%table format
SizeW=[4 6];
varTypesW={'double','double','double','double','double','double'};
varsW={'W(X0)','W(X1)','W(X2)','W(X3)','W(X4)','W(X5)'};
RowsW={'eqR','eqL','eqB','eqDet'};

vars1=[0.932146,2.218409,0.043501,1.028071,0.134353]; %cusppt from matcont
[newDW,newN,N,Wx,DWd,DWw]=NewtonsMethodfcn(eqns,vars1,vars,5);
insertT=Wx(:,1:6);
WX=table('Size',SizeW,'VariableTypes',varTypesW,'VariableNames',
    varsW,'RowNames',RowsW);
WX(:,:)=array2table(insertT)
  
\end{verbatim}

Figure 4
\begin{verbatim}
s=.02;
S=20;
c=200;
m=2;
b=3;
r=5;
a=1000;
syms R L B rho K
assume([R L B], 'Real')
assume([R L B], 'Positive')
eqn1=(1/(1+B))-rho*R;
eqn2=((a*R^r)/(m^r+R^r))-L-c*L*B;
eqn3=s+(((S-s)*K^b)/(K^b+B^b))-B-c*L*B;
eqn4=(-rho)*((-B*c-1)*(-L*c-(B^(b-1)*K^b*b*(S-s))/(B^b+K^b)^2-1)
    -(-L*c)*(-B*c))+(-1/(B+1)^2)*((R^(r-1)*a*r)/(R^r+m^r)-(R^r*R^(r-
    1)*a*r)/(R^r + m^r)^2)*(-B*c);
eqns=[eqn1;eqn2;eqn3;eqn4];
vars=[R,L,B,rho,K];
rhoKFW=[]; %list to save curve coordinates in forward direction
rhoKBW=[]; %list to save curve coordinates in backward direction
vars1=[0.932146,2.218409,0.043501,1.028071,0.134353]; %cusppt from matcont
[newDW,newN,N,Wx,DWd,DWw]=NewtonsMethodfcn(eqns,vars1,vars,5);
rhoKFW=[N(4:5,end)];
rhoKBW=[N(4:5,end)];
    
%adjust index range to plot more points
for k=1:5
        deg=10^(-k);
        ker=null(newDW)*(deg);
        null(newDW);
        x0p=newN+ker;
        x0m=newN-ker;
        x1p=x0p;
        x1m=x0m;
        
        for j=1:20 
            [newDW2,newN2,N2,Wx2,DWd2,DWw2]=
                NewtonsMethodfcn(eqns,x1p',vars,1);
            [newDW3,newN3,N3,Wx3,DWd3,DWw3]=
                NewtonsMethodfcn(eqns,x1m',vars,1);
            
            ker2=null(newDW2)*deg;
            x1p=newN2+ker2;            
            rhoKFW=[rhoKFW,N2(4:5,end)];
            
            ker3=null(newDW3)*deg;
            x1m=newN3-ker3;
            rhoKBW=[rhoKBW,N3(4:5,end)];

        end
 end

%plot(rhoKFW(2,:),rhoKFW(1,:),'o',rhoKBW(2,:),rhoKBW(1,:),'o',0.134353,1.028071,'d')
plot(rhoKFW(2,1:11),rhoKFW(1,1:11),'-', rhoKBW(2,1:11), rhoKBW(1,1:11), '-', 0.134353,1.028071, 'd')
xlabel('K')
ylabel('\rho')
\end{verbatim}

Figure 5
\begin{verbatim}
s=.02;
S=20;
c=200;
m=2;
b=3;
r=5;
p=10;
a=1000;
syms R L B rho K
assume([R L B], 'Real')
assume([R L B], 'Positive')
eqn1=(1/(1+B))-rho*R;
eqn2=((a*R^r)/(m^r+R^r))-L-c*L*B;
eqn3=s+(((S-s)*K^b)/(K^b+B^b))-B-c*L*B;
%rho,K values from MATCONT CONTINUATION
rho0=1.028071; 
K0=0.134353;
D=jacobian([eqn1;eqn2;eqn3],[R,L,B,rho,K]);
vars=[R,L,B,rho,K];

%cusp pt from MATCONT
varscusp=[0.932146,2.218409,0.043501,1.028071,0.134353];

subs(D,vars,varscusp);
null(subs(D,vars,varscusp));
nullsp=null(double(subs(D,vars,varscusp)));
nullsp(4,1);
nullsp(5,1);

vector=[nullsp(4,1)/nullsp(5,1);1];

tvec=linspace(-.01,.01,100);
Bvec=[];

for i=1:50
    rhoK=vector*tvec(i)+[rho0;K0];
    X=vpasolve(subs([eqn1;eqn2;eqn3],[rho,K],[rhoK(1),rhoK(2)]));
    Bvec=[Bvec,X.B];
end

Bvec3=[];
for i=51:100
    rhoK=vector*tvec(i)+[rho0;K0];
    X=vpasolve(subs([eqn1;eqn2;eqn3],[rho,K],[rhoK(1),rhoK(2)]));
    Bvec3=[Bvec3,X.B];
end

T=[tvec(1:50),tvec(51:100),tvec(51:100),tvec(51:100)];
Bt=[Bvec(1:50),Bvec3(1,:),Bvec3(2,:),Bvec3(3,:)];

figure(2)
plot(T,Bt,'.')
xlabel('T')
ylabel('B')
\end{verbatim}

Figure 8

\begin{verbatim}
s=.02;
S=20;
c=200;
m=2;
b=3;
r=5;
a=1000;

syms R L B rho K
assume([R L B rho K], 'Real')
assume([R L B rho K], 'Positive')

eqnB = @(K, rho, B) s + (S-s)*K^b/(K^b+B^b) - B*(1+a*c/((m^r*rho^r*(1+B)^r+1)*(1+c*B)));

fimplicit3(eqnB)
xlabel('K');
ylabel('rho');
zlabel('B');

set(gca, 'Xscale','log')
set(gca, 'Yscale','log')
zlim([0,10])
\end{verbatim}

Supplementary Materials: Figure 12
\begin{verbatim}
syms x y a m
assume([x y a m],'Real')
assume([x y a m],'Positive')

figure
fimplicit3(@(a,m,x) (a/(1+(2/(1+x^m))^m)) - x, [0 3 0 3 0 3])
xlabel('\alpha_1'); ylabel('m'); zlabel('x');
title('2-gene network, plot x vary \alpha_1,m')

figure
fimplicit3(@(a,m,y) (2/(1+(a/(1+y^m))^m)) - y, [0 3 0 3 0 3])
xlabel('\alpha_2'); ylabel('m'); zlabel('y');
title('2-gene network, plot y vary \alpha_2,m')
\end{verbatim}

Supplementary Materials: Figure 13
\begin{verbatim}
alpha=2;
x = -1:.2:3;
y = -1:.2:3;
[X,Y] = meshgrid(x,y);
m = 10; %vary m 
V1 = alpha./(1+Y.^m)-X
V2 = alpha./(1+X.^m)-Y
uV1=V1./sqrt(V1.^2+V2.^2);
uV2=V2./sqrt(V1.^2+V2.^2);
hold on
quiver(X,Y,uV1,uV2,.5,'c')
plot(1,1,"*r",2, 0,"ro",0, 2,"ro",x,x,'r')
str={'saddle',{'sink','high CDK2','low Rb'},{'sink','low CDK2','high Rb'}}
text([1.1 2.1 0.1],[1 0 2],str)
text(-.5,-.5,'x=y','Color','red')
xlabel('x')
ylabel('y')
title('2-gene vector field, m=10, \alpha=2')
hold off
\end{verbatim}

Supplementary Materials: Figures 14 - 16
\begin{verbatim}

alpha=2;
m=3;

%% Solution curves and eigenspace

figure(1)
vec = linspace(0,4,3);
plot3(1,1,1,'k*')
hold on

syms x y z

%eigenvectors of the saddle
x0=[1,1,1] % saddle pt
DxV=jacobian([
    (alpha/(1+z^m))-x;
    ((alpha*x^m)/(1+x^m))-y;
    (alpha/(1+y^m))-z],[x,y,z]);
A=double(subs(DxV,[x,y,z,alpha,m],[x0,alpha,m]));
[evec,eval]=eig(A);
real1=(evec(:,1)+evec(:,2))/2;
real2=A*real1;
%eigenspace
xs=linspace(0,2,5);
ys=linspace(0,2,5);
[Xs,Ys]=meshgrid(xs,ys);

a=((real1/norm(real1))'+x0)-x0;
b=((real2/norm(real2))'+x0)-x0;
n=cross(a,b)
Z=(-n(1).*(Xs-1)-n(2).*(Ys-1))./n(3) + 1;

surf(Xs,Ys,Z,'FaceColor',[1 0 0],'EdgeColor','none','FaceAlpha',.4)
eigdir=[((real1/norm(real1))'+x0);x0;((real2/norm(real2))'+x0)];
plot3(eigdir(:,1),eigdir(:,2),eigdir(:,3),'m-','LineWidth',2)

clear x y z
t=[0,20];
for i=1:length(vec)
    for j=1:length(vec)
        for k=1:length(vec)
            X0=[vec(i),vec(j),vec(k)];
            [t,x_out]=ode45(@xdot, t, X0);
            plot3(x_out(:,1),x_out(:,2),x_out(:,3),'b:')
            plot3(x_out(end,1),x_out(end,2),x_out(end,3),'k*')
        end
    end
end

[t,x_out]=ode45(@xdot, [0 100], [0.5 0.5 2.5]);
sink1=x_out(end,:);
[t,x_out]=ode45(@xdot, [0 100], [2.5,2.5,.5]);
sink2=x_out(end,:)-[0,0,.05];

grid on
xlim([0 2])
ylim([0 2])
zlim([0 2])
xlabel('x')
ylabel('y')
zlabel('z')
title('Plot of solution curves divided by the eigenspace')
legend('Equilibria','Eigenspace')
hold off

%% Surfaces

[x,y] = meshgrid(linspace(-1,10));

yy =  (2*x.^m)./(x.^m+1);
zz =  2./(x.^m+1);
xx =  2./(x.^m+1);

figure(2)
plot3(1,1,1,'r*','MarkerSize',12)
hold on
surf(xx,y,x,'FaceAlpha',0.2,'FaceColor','b','LineStyle','none')
surf(x,yy,y,'FaceAlpha',0.2,'FaceColor','r','LineStyle','none')
surf(x,y,zz,'FaceAlpha',0.2,'FaceColor','g','LineStyle','none')

plot3(sink1(1),sink1(2),sink1(3),'ko','MarkerSize',12)
plot3(sink2(1),sink2(2),sink2(3),'ko','MarkerSize',12)
%surf(x,yy,xx)
xlim([0 2])
ylim([0 2])
zlim([0 2])
xlabel('x')
ylabel('y')
zlabel('z')
grid on
legend('saddle','x','y','z','sink')
title('3-gene solution surfaces and nullclines')
hold off

%% Vector plot
clear x y z
figure(3)
x = 0:.5:2.5;
y = 0:.5:2.5;
z = 0:.5:3.5;
[X,Y,Z] = meshgrid(x,y,z);
V1 = alpha./(1+Z.^m)-X;
V2 = alpha*X.^m./(1+X.^m)-Y;
V3 = alpha./(1+Y.^m)-Z;
uV1=V1./sqrt(V1.^2+V2.^2+V3.^2);
uV2=V2./sqrt(V1.^2+V2.^2+V3.^2);
uV3=V3./sqrt(V1.^2+V2.^2+V3.^2);
quiver3(X,Y,Z,uV1,uV2,uV3,.3)
hold on
plot3(1,1,1,'r*')
plot3(sink1(1),sink1(2),sink1(3),'ko')
plot3(sink2(1),sink2(2),sink2(3),'ko')
xlim([0 2])
ylim([0 2])
zlim([0 2])
xlabel('x')
ylabel('y')
zlabel('z')
title('3-gene vector field, m=3, \alpha=2')
hold off

%% Projections
clear x y z X Y Z
x = linspace(0,4,20);
y = linspace(0,4,20);
z = linspace(0,4,20);
Y =  (2*x.^m)./(x.^m+1);
Z =  2./(y.^m+1);
X =  2./(z.^m+1);

figure(4)
% XY
nexttile
plot(x,Y)
title('Projection on x-y plane')
xlabel('x')
ylabel('y')
% ZX
nexttile
plot(X,z)
title('Projection on z-x plane')
xlabel('x')
ylabel('z')
% YZ
nexttile
plot(y,Z)
title('Projection on y-z plane')
xlabel('y')
ylabel('z')

%% ODE function

function dx=xdot(t,x)
alpha = 2;
m = 3;
dx=zeros(3,1);
dx(1)=(alpha/(1+x(3)^m))-x(1);
dx(2)=((alpha*x(1)^m)/(1+x(1)^m))-x(2);
dx(3)=(alpha/(1+x(2)^m))-x(3);
end

\end{verbatim}

\newpage

\bibliography{references}

\end{document}